\documentclass[final,1p,times]{elsarticle}

\usepackage{amssymb}
\usepackage{amsmath}

\usepackage{array}
\usepackage{multirow}
\usepackage{graphicx}
\usepackage{subcaption}
\usepackage{makecell}
\usepackage{url}
\usepackage{booktabs}
\usepackage{diagbox}
\usepackage{dirtytalk}
\usepackage{blindtext}
\usepackage{float}
\usepackage[table]{xcolor}
\usepackage{setspace}
\usepackage[breaklinks=true]{hyperref}
\usepackage{breakurl}

\journal{Computers and Chemical Engineering}

\begin{document}

\begin{frontmatter}



\title{Using large language models for solving thermodynamic problems}


\author[1]{Rébecca Loubet}
\author[1]{Pascal Zittlau}
\author[2]{Luisa Vollmer}
\author[1]{Marco Hoffmann}
\author[3]{Sophie Fellenz}
\author[1]{Fabian Jirasek}
\author[2]{Heike Leitte}
\author[1]{Hans Hasse\corref{cor1}}
\ead{hans.hasse@rptu.de}
\cortext[cor1]{Corresponding author}


\affiliation[1]{organization={Laboratory of Engineering Thermodynamics},
            addressline={RPTU Kaiserslautern}, 
            city={Kaiserslautern},
            postcode={67655}, 
            state={RP},
            country={Germany}}
            
\affiliation[2]{organization={Laboratory of Visual Information Analysis},
            addressline={RPTU Kaiserslautern}, 
            city={Kaiserslautern},
            postcode={67655}, 
            state={RP},
            country={Germany}}

\affiliation[3]{organization={Laboratory of Machine Learning},
            addressline={RPTU Kaiserslautern}, 
            city={Kaiserslautern},
            postcode={67655}, 
            state={RP},
            country={Germany}}


\begin{abstract}
Large Language Models (LLMs) have made significant progress in reasoning, demonstrating their capability to generate human-like responses. This study analyzes the problem-solving capabilities of LLMs in the domain of thermodynamics. A benchmark of 22 thermodynamic problems to evaluate LLMs is presented that contains both simple and advanced problems. Five different LLMs are assessed: GPT-3.5, GPT-4, and GPT-4o from OpenAI, Llama 3.1 from Meta, and le Chat from MistralAI. The answers of these LLMs were evaluated by trained human experts, following a methodology akin to the grading of academic exam responses. The scores and the consistency of the answers are discussed, together with the analytical skills of the LLMs. Both strengths and weaknesses of the LLMs become evident. They generally yield good results for the simple problems, but also limitations become clear: The LLMs do not provide consistent results, they often fail to fully comprehend the context and make wrong assumptions. Given the complexity and domain-specific nature of the problems, the statistical language modeling approach of the LLMs struggles with the accurate interpretation and the required reasoning. The present results highlight the need for more systematic integration of thermodynamic knowledge with LLMs, for example, by using knowledge-based methods.

\end{abstract}




\end{frontmatter}



\section{Introduction}
\label{sec1}
Thermodynamics is an essential part of both science and engineering, and it is key to understanding energy systems and mitigating climate change. It involves some of the most fundamental principles of physics, such as the first and second laws of thermodynamics, and its applications span almost every field of scientific inquiry. However, while the theoretical foundation of thermodynamics is well-established, applying this knowledge in practice presents a significant challenge. Solving practical thermodynamic problems often involves not only interpreting complex information but also understanding how to adapt and apply these theoretical physical principles to specific scenarios. Currently, most of this work has to be done by human experts, eventually assisted by computational tools, which are, for the most part, designed to solve domain-specific problems. For example, all process simulation tools used in the chemical industry contain hard-coded thermodynamic knowledge. Given the importance of thermodynamics, it is not surprising that its knowledge has also been included in ontologies that were developed for a variety of purposes \cite{PhysSys, OntoCape, OntoKin, propnet, OSMO, PSO, QAChemistry, DigitalTwin}. \par

We have recently developed a system for solving thermodynamic problems, called KnowTD, that is based on such an ontology \cite{KnowTD}. This knowledge-based system selects the equations that describe a given problem from the ontology and then automatically and accurately solves them. KnowTD currently encompasses only a small part of the thermodynamic knowledge, but it is extendable and demonstrates the feasibility of knowledge-based approaches for flexibly solving thermodynamic problems. Another knowledge-based system is WolframAlpha, a computational engine based on the language Mathematica \cite{Mathematica}
that uses symbolic computation and real-time data processing for problem solving and can handle natural language input. It includes some thermodynamic knowledge but can handle only thermodynamic problems from narrow, well-defined classes. Its strength lies in solving the pertinent equations rather than in finding them. \par
\vspace{2mm}

A different approach to solving thermodynamic problems has become practically feasible only very recently: Thermodynamic problems can be given as prompts to large language models (LLMs) in the hope of obtaining a correct answer. This approach is explored in the present work. \par

LLMs are sophisticated computational models designed to process and generate human-like text \cite{Yao_2024}. They operate by converting text into smaller units called tokens, which are numerical representations of words, subwords, or characters. The models are based on a transformer architecture, which uses self-attention mechanisms to capture relationships and dependencies between tokens across an input sequence \cite{vaswani2017attention}. Through training on large datasets, these models learn to predict the likelihood of the next token given the context, enabling them to perform various tasks such as text generation, summarization, and question answering. Their performance is driven by two primary factors: the size of the training corpus, which influences the diversity of linguistic patterns they can learn, and the number of model parameters, which determine their capacity to encode and process complex relationships.\par

LLMs have advanced significantly in recent years, with several new models launched in 2024, including OpenAI's GPT-4o \cite{openai2024gpt}, Meta's Llama 3.1 \cite{llama_meta}, and Mistral's le Chat \cite{MistralAI_Website}. Their applications, once centered on language tasks, have rapidly broadened. This includes the domain of problem solving in science and engineering. In the recent past, a number of studies have been published in which LLMs were used in this domain. These studies often have an educational background \cite{Tsai_2023,wardat2023chatgpt,liang_exploring_2023,krupp2024unreflected,Wang_2024_pitfallsOfChatGPT,Kortemeyer_2024_GradingAssistance,Polverini_2024_UnderstandingGraphsKinematics,Yeadon_2024_GCSE_universityLevels,Sperling_2024_AI_HighScoolPhysics}, and many of them come from the field of mathematics \cite{wardat2023chatgpt,plevris2023chatbots,frieder2024Mathematical,zhang_gap_2020,patel-etal-2021-nlp,lu-etal-2023-survey}. A mixed picture emerges from these studies: While in certain applications impressive strengths of the LLMs in problem solving were found, major drawbacks have also become clear: LLMs are by nature statistical models and so are their answers, i.e., prompting the same question generally leads to different answers that are often of significantly differing quality. Furthermore, prompt design greatly affects the quality of the LLMs' results, which often sound convincing but may be hallucinations \cite{plevris2023chatbots}. Thus, verifying LLM outputs is essential \cite{wardat2023chatgpt}. This need for scrutiny ties into a broader and ongoing debate about the analytical and reasoning skills of LLMs: It is often argued that they lack such skills, e.g., when it comes to graduate-level problems \cite{frieder2024Mathematical}. However, there are also more differentiated views, such as that of \textit{Venkatasubramanian} \cite{venkatasubramanian2024whatsembeddingroseembedding}, who argues that LLMs have what he describes as ``geometrical'' reasoning skills but lack analytical reasoning capabilities. Recently, \textit{Yao et al.} \cite{yao2024tree} have shown that step-by-step solutions, known as a "tree-of-thought", can be used to improve the problem-solving capacity of LLMs. 
A concern with LLMs is their opacity. Some of the most powerful LLMs are essentially black-box models, i.e., the publicly available information on their training and mechanisms is limited \cite{Zhao_2024_LLM_Blackbox}. Hence, empirical testing is the only practical approach to exploring the abilities of LLMs for problem solving. This is what we have done in this study for thermodynamics. \par


Previous studies have used LLMs for problem solving in physics, sometimes also including problems from thermodynamics: \textit{Pursani et al.} \cite{Pursnani_2023_engineeringExamUSFundamentals} evaluated GPT-3.5-Legacy, GPT-3.5-Turbo, and GPT-4 on the 110 problems from the US fundamentals of engineering environmental exam. \textit{K. Wang et al.} \cite{Wang_2024_pitfallsOfChatGPT} used GPT-4 for 40 problems from a college-level engineering physics course, and \textit{X. Wang et al.} \cite{Wang_2023_SciBench} introduce SciBench, a benchmark of 869 problems in mathematics, chemistry, and physics collected from college-level textbooks, and apply different LLMs for their solution. Other work in the field loosely related to thermodynamics is described in \cite{Tsai_2023,Song_01.06.2024,Li2024_simulatingexperts}.\par

Our study complements and extends this in different ways: We focus on a single field, thermodynamics, which is studied in a systematic way. Within this field, we carry out tests with two problem sets of varying complexity and with different LLMs and compare the results. Furthermore, our problems were designed especially for this study, aiming to ensure clarity, consistency, and variety. The problems resemble textbook problems and are given entirely in text form. They are rigorously defined to ensure that each problem has a single, well-determined solution. We provide full access to the problems, solutions, and scoring system used in the Supporting Information, so that they can be reused in future studies, including, eventually, the training of LLMs. Without employing fine-tuning, we conducted preliminary tests using single-shot queries to identify the most effective prompting strategy. This process involved evaluating various general instructions and ways of presenting the problems. The optimal prompting strategy, determined through these tests, was consistently applied for the studies on which we report here. All prompts utilized in this research are disclosed herein. To assess the variance of the answers, each problem was given to each studied LLM three times in independent queries. The answers were then graded by trained domain experts based on a standard solution. Interestingly, problems from thermodynamics have been used very recently as test cases to study whether LLMs can be applied to assist in grading exams \cite{Kortemeyer_2024_GradingAssistance}. In a nutshell, the answer is: No.\par
\vspace{2mm}

The research questions we address in our work are: 
\begin{enumerate}
  \item How good are LLMs at solving thermodynamic problems?
  \item How does the performance vary with the level of complexity of the problems?
  \item How do different state-of-the-art LLMs perform on the same test sets?
  \item What changes occur when going from one to the next version of the LLM?
  \item What is the variance in the answers when the same prompt is repeated?
  \item What types of errors occur?
\end{enumerate}
The results of our study contribute to the important ongoing discussions of the problem-solving capabilities of LLMs within the scientific and engineering domains. Furthermore, we compare the problem-solving performance of LLMs with that of knowledge-based systems, specifically KnowTD \cite{KnowTD}. While this study does not explicitly address educational issues, the outcomes may also offer insights relevant to that field.\par
\vspace{2mm}

This paper is organized as follows: The studied LLMs and how they were used are described in Section~\ref{LLMs}. In Section~\ref{problems}, we present the problems that were classified in two sets, a simple and an advanced one. The scoring method is described in Section~\ref{evaluation}. The results obtained with the different LLMs on the different test sets are presented and discussed in Section~\ref{results} and compared to those of KnowTD in Section~\ref{KnowTD}. We conclude in Section~\ref{conclusion}.\par



\section{Studied LLMs and their Utilization}
\label{LLMs}
In this study, we have tested the following LLMs: GPT-3.5, GPT-4, and GPT-4o from \mbox{OpenAI}, Llama 3.1 from Meta, and le Chat from MistralAI. The details of these LLMs are listed in Table~\ref{table:LLM Overview}. While the inner workings of GPT models are undisclosed, the code and parameters for Llama 3.1 and le Chat are publicly available and can be accessed through platforms like HuggingFace. Therefore, Llama and le Chat are often referred to as open-source models. However, neither the training procedure nor the training data for these models are publicly available. In particular, we do not have specific information regarding the extent to which thermodynamic knowledge was incorporated into the training of any of the studied models. OpenAI discloses the cutoff dates for its models' training: September 2021 for GPT-3.5, April 2023 for GPT-4, and October 2023 for GPT-4o \cite{OpenAI_Website_3.5, OpenAI_Website_4, OpenAI_Website_4o}. Meta reports that Llama 3.1 was trained on $15\cdot10^{12}$ multilingual tokens \cite{vavekanand2024llama}, while limited information is available regarding the training of le Chat \cite{MistralAI_Dataset}. The context window of LLMs in Table~\ref{table:LLM Overview} refers to the maximum number of tokens the model can process simultaneously.
\par

\setlength{\tabcolsep}{4pt}
\begin{table}[h!]
\caption{Overview of the LLMs studied in this work: The references are either sources or indicate additional information.}
\centering
\begin{tabular}{|l|c|c|c|c|c|c|}
 \hline
 LLM &  Company & \makecell{Code \\ type} & Release & \makecell{Context \\ window \\ tokens} & \makecell{Parameters \\ / $10^9$} & \makecell{ Python \\ code \\ interpreter} \\ [0.5ex] 
 \hline
 GPT-3.5 \cite{OpenAI_Website_3.5} & OpenAI & Proprietary & 11.2022 \cite{xu2024surveymultilinguallargelanguage} & 16,385 & \makecell{175 \\ \cite{Rudolph_2023,xu2024surveymultilinguallargelanguage, Haque_2024}} & No \\ 
 \hline
 GPT-4 \cite{OpenAI_Website_4} & OpenAI & Proprietary & 03.2023 \cite{Zaitsu_2023} & 128,000 &  \makecell{$\sim$ 1,760 \\ \cite{Haque_2024,xu2024surveymultilinguallargelanguage}} & Yes \\ 
 \hline
 GPT-4o \cite{OpenAI_Website_4o} & OpenAI & Proprietary & 05.2024 \cite{Shahriar_2024} & 128,000 & \makecell{$>>$ 1,000 \\ \cite{Shahriar_2024}} & Yes \\
 \hline
 Llama 3.1 \cite{llama_meta} & Meta & \makecell{Open \\ source} & 07.2024 \cite{llama_meta} & 128,000 & 70 \cite{llama_meta} & No \\
 \hline
 \makecell[l]{Le Chat \cite{MistralAI_Website} \\ (Mistral Large 2)} & MistralAI & \makecell{Open \\ source} & 07.2024 \cite{MistralAI_Website} & 128,000 & 123 \cite{MistralAI_Website} & No \\ [1ex]
 \hline
\end{tabular}
\label{table:LLM Overview}
\end{table}

The prompts were generated consistently for all LLMs. The idea behind prompting was to think of the LLM as a student in an exam: A single-shot query approach was employed, where no feedback, intermediate guidance, or refinement was provided. Additionally, the LLMs were not fine-tuned prior to the problem-solving tasks. Each prompt was introduced with the following header: \par 
\begin{quote}
\textit{"You are a student in a thermodynamics class. You are given problems and must find the solution. The answer should be formatted in the same way as the input to be readable in a standard text editor; do NOT use LaTeX encoding. Write formulas like this: p\_1*v\_1 = R*T\_1. Write units like this: J/kg. Use a code generator."}
\end{quote}
Since le Chat and Llama 3.1 had the tendency to give long and verbose answers, the prompt was modified for these LLMs by adding:\par
\begin{quote}
\textit{"Please answer in a concise, scientific form."}
\end{quote}
Since the responses of LLMs to a given prompt are stochastic, i.e., the same prompt can yield different answers when submitted multiple times, each problem was presented three times to each of the studied LLMs. These repetitions were not part of the same dialogue and can, therefore, be considered to be independent; the LLMs did not retain information or learn from one repetition to the next. This approach represents a balance between the desire to increase the number of trials to enhance statistical reliability and the practical constraints of manually evaluating and correcting each response. \par
We have accessed the different models via APIs, except for Llama 3.1, which was downloaded and run locally on a GPU. Therefore, we were able to work with the largest available models at the time of the study, with the exception of Llama 3.1: While the largest model has $405\cdot10^9$ parameters, our work was conducted with the $70\cdot10^9$ parameter-model. \par

Several parameters control how LLMs generate their responses. In the following, we introduce the key parameters and report the values used for each LLM, which are the default values in all cases. The \textit{temperature} parameter regulates the creativity and variability of the model's outputs. Lower \textit{temperature} values (i.e., below 1.0) lead to more deterministic and repetitive responses, while higher values (i.e., above 1.0) encourage greater creativity and randomness. For the models in this study, the \textit{temperature} was set to 1.0 for GPT models, 0.8 for Llama 3.1, and 0.7 for le Chat.
Another key parameter is the \textit{top\_p} value, which influences the diversity and quality of the model's responses by limiting the selection of tokens to those within a specified cumulative probability mass. A higher \textit{top\_p} value (e.g., 1.0) allows a broader range of tokens, increasing diversity, while lower values narrow the selection. For this study, \textit{top\_p} was set to 1.0 for both the GPT models and le Chat, and 0.9 for Llama 3.1.
The \textit{presence\_penalty} parameter encourages or discourages the introduction of new topics or tokens. Positive values penalize the introduction of new tokens based on their prior occurrence in the text, while negative values favor staying on previously discussed topics. The default setting for \textit{presence\_penalty} was 0 for all five LLMs.

\section{Benchmark Problems and Evaluation}
\label{problems}

\subsection{Overview}
\label{overview LLMs}
In this study, two sets of thermodynamic benchmark problems were used, which we will refer to as simple problems and advanced problems, respectively. Both sets include problems resembling those typically encountered in an introductory university course in engineering thermodynamics. Solving these problems requires the combination of different elements of thermodynamics. For a trained human, the simple problems are relatively straightforward -- much easier than what we would require in an introductory course exam. The advanced problems approach the exam level but do not reach it. \par

All problems are reported in electronic form online\footnote{https://gitlab.rhrk.uni-kl.de/knowtd/llm-benchmark-testsets} and in the Supporting Information, which also includes information on their solutions and the maximum scores that could be reached. The problems were designed to be fully specified by text and have unique solutions. All tasks require the calculation of numerical values of variables. The set of simple problems used here is identical to the set of problems that we have recently used as examples to illustrate the application of KnowTD \cite{KnowTD}.

\subsection{Simple Problems}
\label{simple prb.}
The 13 problems in this set are problems in which closed systems that contain an ideal gas undergo a single change of state from one state of equilibrium to another. The considered changes of state include different types of processes, including: 
\begin{itemize}
  \item Isothermal, isobaric, isochoric, polytropic
  \item Adiabatic, non-adiabatic
  \item Reversible, irreversible.
\end{itemize}
Heat and/or work can be supplied to the system. The problems in the set cover the feasible combinations of the main features, which makes this set different from many other test sets used to evaluate LLMs, in which the problems are basically drawn stochastically from a large pool. If all problems from the present set of simple problems can be solved, we may assume that any other well-posed problem from the considered class of problems can also be solved.  As an example, we present Problem 1 here:\par
\begin{quote}
\textbf{\textit{Problem 1}}\par
\textit{A gas in a cylinder is compressed reversibly from v\_1~=~0.05~m\^{}3/kg to v\_2~=~0.02~m\^{}3/kg. The initial temperature is T\_1~=~298~K. The process is adiabatic. What is the work supplied per kilogram of gas?}
\end{quote}

\subsection{Advanced Problems}
\label{adv.problems}
This set contains nine problems that cover a wider range of thermodynamic knowledge than the first set. It includes processes with closed and open systems, multi-step processes (including cyclic processes), processes with liquids and gases, and processes with coupled systems. Some of these problems also include the contribution of kinetic and/or potential energy. In each problem, several questions are asked, which must typically be solved sequentially. Each of these questions has a maximum score. As an example, we present Problem 18 here:\par
\begin{quote}
\textbf{\textit{Problem 18}}\par
    \textit{In a test rig, two streams of water are taken from two tanks (1 and 2) in which the water is stored at ambient conditions (T\_1 = T\_2 = 300 K, p\_1 = p\_2 = 1 bar). Tank~1 is located at the height of the rig, tank 2 is at a height of z\_2 = 30 m above the rig. The mass flow rates of both streams are equal. The water from tank 1 is isobarically heated in a heat exchanger, the water leaving the heat exchanger in state 3 is fed into a first inlet of an adiabatic mixer. The water from the second tank passes a line leading it to the rig and is then fed into the second inlet of the mixer in state 4. The water that leaves the mixer in state 5 is then fed into an isentropic pump with a mechanical power of 260 W. The water leaving the pump in state 6 is accelerated in an isentropic nozzle. The velocity of the water leaving the nozzle in state 7 is c\_7 = 30 m/s, the pressure is p\_7 = 1 bar, and the temperature is T\_7 = 330 K. The cross-section of the outlet of the nozzle is A\_7 = 0.2 cm\^{}2}.\\
\renewcommand{\labelenumi}{\textit{\alph{enumi})}}
\begin{enumerate}
    \item \textit{Calculate the mass flow rate of the stream leaving the nozzle.}
    \item \textit{Calculate the heat flow in the heat exchanger and the resulting temperature~T\_3.}
    \item \textit{Calculate the pressures p\_5 and p\_6.}
\end{enumerate}\vspace*{10pt}
	
	\textit{Assumptions:}
	\begin{itemize}
		\item \textit{The kinetic energy in states 1 to 6 is negligible.}
		\item \textit{All connecting lines are adiabatic.}
		\item \textit{The process is stationary.}
	\end{itemize}\vspace*{10pt}
	
	\textit{Fluid properties:}
	\begin{itemize}
		\item \textit{Water is considered to be a constant density fluid with rho = 1000 kg/m\^{}3 and c = 4.2 kJ/(kg K).}
		\item \textit{The gravitational acceleration is g = 9.81 m/s\^{}2.}
	\end{itemize}
\end{quote}

\subsection{Evaluation}
\label{evaluation}
The responses of the LLMs were evaluated by trained experts as if the LLMs had been given the problems in an academic exam. Since the task was always to calculate numerical values of certain variables, an option for the evaluation would have been to check whether or not this value was correct. While this has the advantage of being unambiguous (if numerical issues are left apart), it is not fair because a simple miscalculation could lead to a loss of all points despite a completely correct solution path. Furthermore, considering only the numerical correctness would introduce a bias in favor of LLMs that have the capability to use Python scripts for calculations, which only two out of the five studied LLMs can. Finally, in some cases, even a correct numerical value can be found for wrong reasons. Hence, we have decided to use the more tedious method to evaluate the correctness of the solution path, where partially correct solutions also increase the score. As a rule, in our evaluation, 0.5 points were awarded for each correctly executed step towards the final solution. We acknowledge that the scores obtained this way are not unambiguous. To reduce the variability, we have set up a master solution for each problem with suggestions for intermediate scores. However, this cannot entirely solve the issue of ambiguity, as there are usually many ways to obtain a certain solution. Comparing independent evaluations of the same answer by different experienced reviewers, we estimate the uncertainty of the scores to be typically +/- 0.5 points. A technical problem with this evaluation is that it can only be carried out by experienced personnel, that it requires cross-checking of the evaluation by different reviewers for non-trivial cases, and that it is time-consuming, which limits the scope of any empirical study like the one we have conducted.\par

As an example of our evaluation process, we provide in the Supporting Information a graded response for the two problems discussed in Sections~\ref{simple prb.} and~\ref{adv.problems}. These responses are accompanied by colored annotations, highlighting the awarded points (in green) and the identified errors (in red).

\section{Results and Discussion}
\label{results}
We have tested the performance of the five LLMs listed in Table~\ref{table:LLM Overview} using the two sets of benchmark problems described in Section~\ref{problems}. The corresponding prompt design is described in Section~\ref{LLMs}.

\subsection{Performance Evaluation}
\label{performance}
\begin{figure}
    \centering
    \includegraphics[]{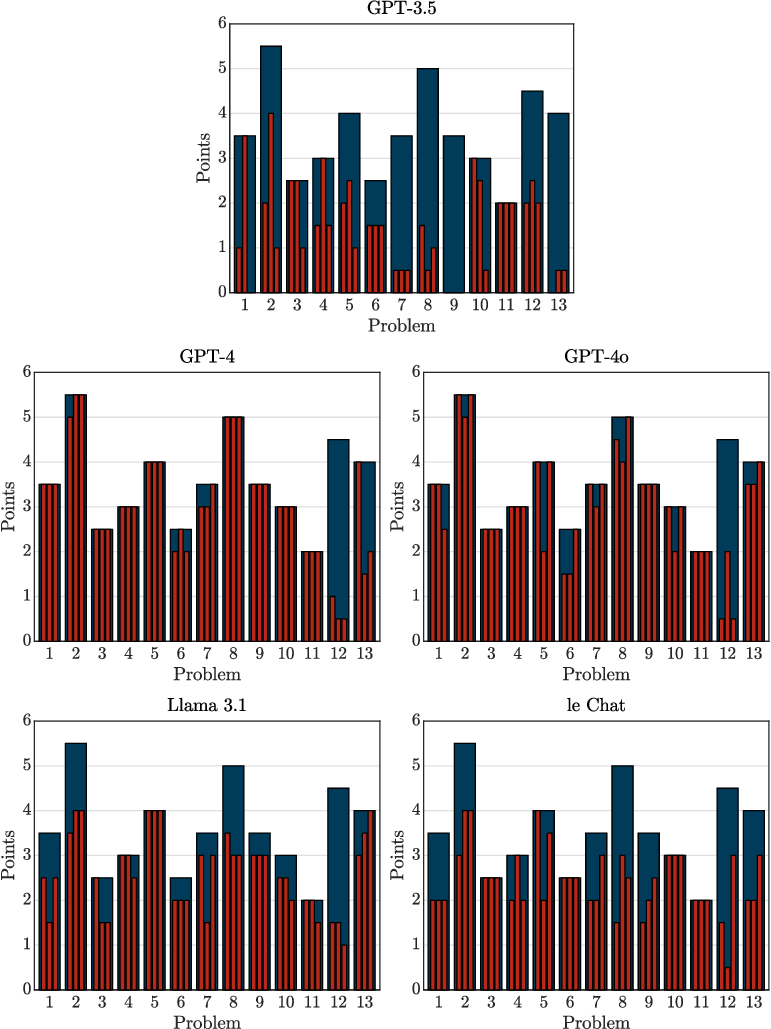}
    \caption{Scores of the different LLMs for the 13 simple problems. Blue bars indicate the maximum number of points, red bars indicate the scores obtained in the three independent runs.}
    \label{fig:plots_simple problems}
\end{figure}

\begin{figure}
    \centering
    \includegraphics[]{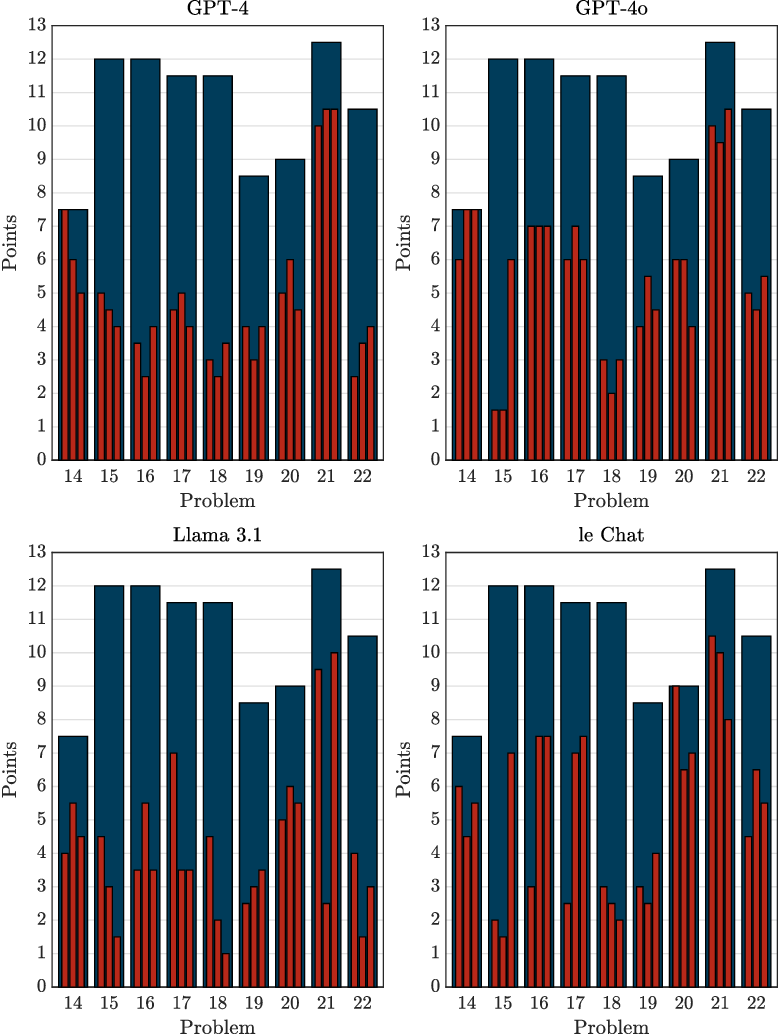}
    \caption{Scores of the different LLMs for the nine advanced problems. Blue bars indicate the maximum number of points, red bars indicate the scores obtained in the three independent runs.}
    \label{fig:plots_advanced problems}
\end{figure}

Figure~\ref{fig:plots_simple problems} gives an overview of the scores of the different LLMs for the simple problems. Figure~\ref{fig:plots_advanced problems} shows the corresponding results for the advanced problems. For each problem, the results from the three independent runs are shown, which allows us to assess the variance of the answers of the different LLMs. As GPT-3.5 consistently failed to generate meaningful responses for the advanced problems, we have refrained from including its results in this study. In some cases, it even terminated with an error. This task failure illustrates the limitations of GPT-3.5 when confronted with complex problems.

Overall, the LLMs perform well on the simple problems (see Figure~\ref{fig:plots_simple problems}). There is a substantial improvement going from GPT-3.5 to GPT-4, but no significant performance differences between GPT-4 and GPT-4o. The results of Llama 3.1 and le Chat are overall similar, with neither model excelling nor completely failing, but these LLMs generally perform worse than GPT-4 and \mbox{GPT-4o}. The scores also show the variability of the responses of all models: Neither of the five LLMs consistently produces the same results when solving the same problem independently multiple times. This becomes particularly evident for problems that were only partially solved. \par

The LLMs struggle particularly with Problem 12, which involves an adiabatic process. The text does not specify if the process is reversible or irreversible, so assuming reversibility is incorrect, yet each studied LLM made this assumption in all three repetitions.\par
\vspace{2mm}

The performance of the LLMs on the advanced problems is shown in Figure~\ref{fig:plots_advanced problems}. As expected, the performance of all LLMs on the advanced problems is significantly worse than on the simple problems. Correct solutions were obtained only in a few cases and never in all three repetitions. Overall, GPT-4o generated the best results; on the advanced problems, it also performed better than GPT-4 (in contrast to what was found for the simple problems). The results of le Chat, which performed worse than GPT-4o on the simple problems, are almost as good as those from GPT-4o for the advanced problems. The variance of the scores for repeated questions is lowest for the GPT models but of similar magnitude for all models. \par
\vspace{2mm}

Table~\ref{table: Statistics_results} presents the results of the statistical evaluation of the scores achieved by the different LLMs on both the simple and advanced benchmark tests. For this purpose, the individual scores on the different problems were normalized with the maximum number of points for each problem (completely correct solution). The drop in the average scores for all models when transitioning from the simple to the advanced problems is clearly observed, along with the performance differences between the various LLMs, as discussed earlier. The statistical evaluation also gives quantitative measures of the considerable variance in the answers of the LLMs. The standard deviation typically ranges between 20 – 25\% of the maximum achievable score, corresponding to 30 – 50\% of the actual scores obtained. This indicates not only that the models fail to produce consistent results for repeated questions (see Figures~\ref{fig:plots_simple problems} and~\ref{fig:plots_advanced problems}), but also that their performance quality varies significantly across different problems. \par

\setlength{\tabcolsep}{5.5pt}
\setlength{\arrayrulewidth}{0.5pt} 
\begin{table}[h!]
\caption{Statistical evaluation of the scores of the different studied LLMs in the simple and the advanced benchmark tests: Mean values $\bar{x}$ and standard deviations $\sigma$. The numbers were calculated after normalizing the scores with the maximum number of points for each problem.}
\centering
\renewcommand{\arraystretch}{1.3} 
\begin{tabular}{|c|c|c|c|c|c|c|}
\hline
\multirow{2}{*}{\centering LLM} & \multicolumn{2}{c|}{Simple Problems} & \multicolumn{2}{c|}{Advanced Problems} \\
\cline{2-5}
 & $\bar{x}$ / \% & $\sigma$ / \% & $\bar{x}$ / \% & $\sigma$ / \%\\
\hline
GPT-3.5 & 45.79 & 34.72 & - & -  \\
GPT-4 & 88.57 & 25.04 & 47.52 & 21.59 \\
GPT-4o & 87.09 & 23.27 & 55.19 & 22.89 \\
Llama 3.1 & 75.88 & 20.31 & 40.70 & 19.88 \\
Le Chat & 72.82 & 24.28 & 51.93 & 23.81 \\
\hline
\end{tabular}
\label{table: Statistics_results}
\end{table}


\subsection{Error Types}
\label{Error Types}
To better understand the nature of the errors of the LLMs, we have classified them into six categories:
\begin{itemize}
    \item \textbf{Wrong assumptions}: The LLM makes an invalid assumption that is not aligned with the problem.
    \item \textbf{Numerical errors}: The LLM makes mistakes in arithmetic operations, resulting in incorrect numerical answers.
    \item \textbf{Inconsistent signs}: The LLM incorrectly applies positive and negative signs in calculations, leading to errors such as incorrect directions of heat flows or forces.
    \item \textbf{Wrong equations}: The LLM selects or derives incorrect equations to solve the given problem, leading to false results.
    \item \textbf{Context confusion}: The LLM fails to capture or retain the necessary context, leading to incorrect solution paths or a complete loss of direction.
    \item \textbf{Verbose aberration}: The LLM gives verbose answers with excessive unnecessary details and thereby loses clarity and focus, ending up in being unable to produce any meaningful solution.
\end{itemize}

The errors were classified by the experts who corrected the answers. An overview of the classification of the errors observed in the benchmark tests is presented in Table~\ref{table: error class_simple problems} for the simple problems and in Table~\ref{table: error class_adv problems} for the advanced problems. The number count in these tables was determined as follows: For a problem with a maximum score of \( N_{\text{max}} \) points, if an LLM obtained \( N \) points, the difference \( N_{\text{max}} - N \) was divided into increments of 0.5 points. Each increment is referred to as a penalty, which was then assigned to one of the error classes listed above.  In case of conflicts in the error class assignment, the class that was deemed to be dominant was chosen,  or, when a given error led to more than one penalty, the penalties were split up between the pertinent classes. There is obviously ambiguity in this process and the numerical results should not be over-interpreted, but they still give an impression of how often a given type of error occurred. \par

\setlength{\tabcolsep}{2pt} 
\renewcommand{\arraystretch}{1.3} 
\begin{table}[h!]
\caption{Survey of the occurrence of errors from the different classes in the benchmark tests on the simple problems. The numbers are the penalties. A penalty corresponds to 0.5 points that were not given due to the corresponding error (see text).}
\centering
\begin{tabular}{l>{\centering\arraybackslash}p{1.5cm}>{\centering\arraybackslash}p{1.5cm}>{\centering\arraybackslash}p{1.5cm}>{\centering\arraybackslash}p{1.5cm}>{\centering\arraybackslash}p{1.5cm}>{\centering\arraybackslash}p{1.5cm}|>{\centering\arraybackslash}p{1.5cm}}
\toprule
\textbf{LLM} & \rotatebox{75}{\textbf{Wrong assumptions}} & \rotatebox{75}{\textbf{Numerical errors}} & \rotatebox{75}{\textbf{Inconsistent signs}} & \rotatebox{75}{\textbf{Wrong equations}} & \rotatebox{75}{\textbf{Context confusions}} & \rotatebox{75}{\textbf{Verbose aberration}} & \textbf{Total} \\
\midrule
\textbf{GPT-3.5} & 14 & 19 & 7 & 12 & 90 & 22 & 164 \\
\textbf{GPT-4}   & 23 & 2 & 3 & 6 & 0 & 3 & 37 \\
\textbf{GPT-4o}   & 21 & 8 & 6 & 5 & 0 & 0 & 40 \\
\textbf{Llama 3.1}   & 12 & 23 & 5 & 27 & 1 & 5 & 73 \\
\textbf{le Chat}     & 15 & 47 & 4 & 14 & 8 & 0 & 88 \\
\bottomrule
\end{tabular}
\label{table: error class_simple problems}
\end{table}

\setlength{\tabcolsep}{2pt} 
\renewcommand{\arraystretch}{1.3} 
\begin{table}[h!]
\caption{Survey of the occurrence of errors from the different classes in the benchmark tests on the advanced problems. The numbers are the penalties. A penalty corresponds to 0.5 points that were not given due to the corresponding error (see text).}
\centering
\begin{tabular}{l>{\centering\arraybackslash}p{1.5cm}>{\centering\arraybackslash}p{1.5cm}>{\centering\arraybackslash}p{1.5cm}>{\centering\arraybackslash}p{1.5cm}>{\centering\arraybackslash}p{1.5cm}>{\centering\arraybackslash}p{1.5cm}|>{\centering\arraybackslash}p{1.5cm}}
\toprule
\textbf{LLM} & \rotatebox{75}{\textbf{Wrong assumptions}} & \rotatebox{75}{\textbf{Numerical errors}} & \rotatebox{75}{\textbf{Inconsistent signs}} & \rotatebox{75}{\textbf{Wrong equations}} & \rotatebox{75}{\textbf{Context confusions}} & \rotatebox{75}{\textbf{Verbose aberration}} & \textbf{Total} \\
\midrule
\textbf{GPT-4}   & 63 & 5 & 0 & 18 & 201 & 19 & 306 \\
\textbf{GPT-4o}  & 71 & 1 & 1 & 27 & 156 & 8 & 264 \\
\textbf{Llama 3.1}   & 93 & 23 & 0 & 27 & 165 & 36 & 344 \\
\textbf{Le Chat}     & 109 & 31 & 0 & 50 & 87 & 1 & 278\\
\bottomrule
\end{tabular}
\label{table: error class_adv problems}
\end{table}

Table~\ref{table: error class_simple problems} shows that not only the number of penalties is much higher for GPT-3.5 than for all other models, but also the distribution of the errors on the classes differs: For GPT-3.5, context confusions are most important, which play only a minor role for the other LLMs. Comparing the results for the other models on the simple problems, as shown in Table~\ref{table: error class_simple problems}, it is clear that the GPT models not only have fewer penalties but also exhibit different error patterns. For the GPT models, the primary source of errors is wrong assumptions, while for Llama 3.1 and le Chat, numerical errors and wrong equations are most important. The reason for this is that even though both Llama 3.1 and le Chat generate code as part of their language modeling tasks, they do not have specialized code generation capabilities like GPT-4 and GPT-4o.\par

The picture changes remarkably when going from the simple benchmark to the advanced benchmark: As expected, there is a significant increase in the number of penalties (see Table~\ref{table: error class_adv problems}). However, the deterioration in performance is more important for the GPT models than for \mbox{Llama 3.1} and le Chat, which leads to a leveling of the quality of the results. This also holds for the types of errors
that occurred. For all LLMs, the most important errors are now context confusions and wrong assumptions. While numerical errors continue to be significant for \mbox{Llama 3.1} and le Chat, they are only a minor issue for GPT-4 and GPT-4o, which aligns with the findings from the simple benchmark (see Table~\ref{table: error class_simple problems}). However, if we consider a scenario in which all LLMs were capable of directly producing correct code and applying it, thereby eliminating the influence of numerical errors, the differences between the LLMs would shrink, both for the simple and the advanced problems. \par
\vspace{2mm}

The high importance of context confusions that was found for all LLMs for the advanced problems is somehow similar to what was found for GPT-3.5 for the simple problems. This weakness of GPT-3.5 was overcome by the more recent versions GPT-4 and \mbox{GPT-4o}, which performed well on the simple problems. Whether or not this trend will also hold for future versions of GPT applied to the advanced problems remains to be seen.\par
\vspace{2mm}

We conclude this discussion with a few specific observations: GPT-3.5, GPT-4, and \mbox{Llama 3.1} occasionally rely on fluid properties from undisclosed sources instead of using the provided data, which carries the risk of inducing avoidable errors. Additionally, both GPT-4 and \mbox{Llama 3.1} tend to generate excessive text that essentially paraphrases problem statements without progressing in the analysis. This verbose output often results in overlooking crucial solution steps, leading to superficial and sometimes repetitive answers without obtaining the desired solution. In contrast, the responses from GPT-4o and le Chat tend to be shorter, better structured, and more targeted. The following quote illustrates verbose aberrations:\par\vspace{2mm}

\textit{"However, this result is incorrect because it doesn't take into account the fact that the piston is compressed by an external force. We need to use the definition of work for a non-equilibrium process [...] Wait! This result is still incorrect. Let's think about it again... [...] No! This result is still incorrect. Let's try again..."}
Llama 3.1 - Advanced Problems - Problem 16, b) - Run 2\par\vspace{2mm} 
This quotation is also interesting, as it illustrates a feature that we have observed quite regularly: The LLMs seem to have the ability to reflect their answers and criticize them – sometimes even coming up with a correct solution in the next try. We can only speculate that after having obtained a wrong result, the LLM finds all choices for his next steps so improbable that it goes back and re-iterates previous steps. In particular, GPT-4 and GPT-4o demonstrate this critical approach to their own responses. While they may initially start with incorrect assumptions, they frequently correct themselves to find the right solution path. The GPT-4o model, in particular, tends to stop when it reaches a point where it can no longer solve the problem rather than proceeding with incorrect assumptions. As a result, it makes fewer errors compared to the other models but often lacks the necessary knowledge to fully solve more complex problems.  \par
\vspace{2mm}

Of the simple problems, Problem 12 was the most challenging for all LLMs. They all assumed that the process was reversible, which is not allowed unless explicitly stated. The same difficulty occurs also in the advanced Problems 19 and 22. Interestingly, some of the responses reflected this point:
\par\vspace{2mm} 

\textit{"To find ( T\_2 ), use the adiabatic compression relation (\textbf{assuming isentropic for simplicity}):"}
le Chat - Advanced Problems - Problem 19, a) - Run 1\par\vspace{2mm} 
\par\vspace{2mm} 

\textit{"To calculate the work input to the compressor per kilogram of working fluid, we first need to find T2 using the isentropic relation for a perfect gas because the process 1→2 is adiabatic and irreversible but we can use an isentropic relation \textbf{to get an estimation}"}
Llama 3.1 - Advanced Problems - Problem 22, a) - Run 2\par\vspace{2mm} 

Other common incorrect assumptions included neglecting contributions of the kinetic energy and treating a liquid as an ideal gas – in some cases with some reflection on this fundamental error: \par
\vspace{2mm} 

\textit{"P1V1 = P2V2. Although this law strictly applies to gases, we'll apply it here as a \textbf{simplification for educational purposes}"}
Llama 3.1 - Advanced Problems - Problem 16, b) - \mbox{Run 1}\par\vspace{2mm} 

\subsection{Preliminary tests with OpenAI's o1-preview model}
\label{o1-preview}
In mid-September 2024, OpenAI released the o1-preview model, which OpenAI claims to deliver answers comparable to those of PhD students on challenging benchmark tasks in physics, chemistry, biology, mathematics, and coding, demonstrating substantial improvements over previous models \cite{OpenAI_Website_o1-preview,Lingo_14.09.2024}. Unlike the other GPT models, which tend to provide instant answers without much internal deliberation, OpenAI o1-preview is built to "think" step-by-step using a technique known as "chain of thought reasoning". \textit{Winter et al.} \cite{de2024system} report that this new model achieved a perfect score on a mathematics final exam for high school students. We therefore decided to carry out a preliminary test with the OpenAI o1-preview model on the simple \mbox{Problem 12}, which none of the other LLMs were able to successfully solve. This test aimed to assess whether the advanced reasoning capabilities of OpenAI o1-preview could overcome the challenges that had previously stumped the other models. The prompt was run three times, yielding the following results: 0.5/4.5, 4.5/4.5 and 3/4.5. While OpenAI o1-preview was the first model to produce a correct solution for Problem 12, the inconsistency across the attempts suggests that it cannot yet be relied upon to consistently generate accurate answers. Even the most advanced LLMs still exhibit limitations in understanding and reasoning -- even for simple thermodynamic problems. To achieve higher reliability in problem solving, a more robust and trustworthy approach is obviously needed. The most promising way to achieve this is to integrate explicit knowledge by combining the LLMs with a knowledge-based system.

\section{Knowledge Based Systems for Thermodynamic Problem Solving}
\label{KnowTD}
While LLMs have demonstrated significant potential for solving problems in diverse domains, the quality of their responses varies strongly, and they struggle in domains that require rigorous application of formal knowledge, such as thermodynamics.
The integration of LLMs and knowledge-based systems is a promising way to address these problems by exploiting the strengths of both approaches. LLMs excel at processing and interpreting large amounts of unstructured data, while knowledge-based systems offer structured, rule-driven approaches grounded in expert understanding. Combining the flexibility and adaptability of LLMs with the precision of knowledge-based methods is therefore highly attractive for solving complex scientific and engineering problems, ensuring both scientific rigor and interpretability in the solutions provided.

Before we discuss ways to intertwine both approaches in the field of thermodynamics, we first provide a brief overview of KnowTD, a knowledge-based system recently developed for solving thermodynamic problems, which will serve as our example. 
KnowTD~\cite{KnowTD} is a specialized knowledge representation system designed to address thermodynamic problem-solving. It is built around an ontology that encodes thermodynamic theories, material properties, and process-related knowledge. This ontology is paired with a reasoner that utilizes the structured knowledge to define, set up, and solve user-specified problems. The ontology was meticulously curated by domain experts to ensure that it contains accurate and validated thermodynamic knowledge. It organizes thermodynamic theory into concepts, attributes, variables, and equations,  along with their interconnections and dependencies. Additionally, it contains information on how these elements can be combined to represent given processes.

Currently, only simple problems can be solved by KnowTD, however, KnowTD is a modular system that can be expanded to include additional knowledge in the future. The reasoner in KnowTD first helps the user enter known and desired information, using the ontology to supplement the problem with related facts. It then constructs a knowledge graph containing all relevant variables and equations applicable to the specified problem and analyzes this graph, tracing a path from the given inputs to the target variables. This approach not only provides a numerical solution but also generates a transparent step-by-step solution path –- a guarantee for correctness. \par

A key feature of KnowTD is its ability to generate traceable and explainable solutions, making it valuable for both problem solving and educational purposes.  The knowledge represented in the current version of KnowTD enables solving all problems from the simple test set presented in Section~\ref{simple prb.}. However, in its current form, KnowTD cannot solve the advanced problems from the second set presented in Section~\ref{adv.problems} -- simply because they are outside the scope of the thermodynamic knowledge that is currently implemented in KnowTD.

The following section briefly discusses key research questions regarding the hybridization of LLMs and knowledge-based systems, using KnowTD as an example of the latter. There are two primary approaches to hybridizing these systems: First, using LLMs to enhance the performance of the knowledge-based system, and second, leveraging the knowledge-based system to improve the performance of the LLM.

\paragraph{Using LLMs to enhance the performance of KnowTD}
The input to KnowTD is a problem posed in natural language, which must be translated into a machine-readable format by the user through a guided dialogue with the system. This task primarily requires text comprehension skills and basic thermodynamic knowledge to interpret the text. LLMs could facilitate this process by acting as an intermediary, converting the problem text into the structured input needed by KnowTD, with the ontology serving as the target "language." The problem text typically refers to real-world items and processes, while the ontology includes the theoretical elements necessary to construct the corresponding mathematical model. Each of the theoretical elements of the ontology may correspond to a wide range of items in the real world. For example, a closed system may represent a tank filled with biogas, liquid in a syringe, or gas in the cylinder of a compressor. LLMs could assist or even replace users in mapping these real-world items to the corresponding theoretical elements within the ontology.

Building on this idea, LLMs could also be used to translate the output of KnowTD into natural language. The results in Section~\ref{Error Types} show that, although LLMs may introduce errors in their reasoning, they are still capable of providing clear and coherent textual explanations of their answers. By translating the graph-based solution path generated by KnowTD into natural language, LLMs could enhance the usability of KnowTD, making it more accessible and intuitive for users.

LLMs might also play a role in the reasoning process of KnowTD. Based on the ontology, KnowTD provides a set of valid equations that describe the problem. However, this set generally contains many more equations than needed to solve the problem. It is a difficult problem to reduce this large, over-defined set of nonlinear equations to a set of equations that enable the solution of the problem without containing redundant equations –- not to mention finding the set of equations that enables the most efficient solution. The pertinent equations and their associated variables can be represented as a graph. Currently, single-equation traversal is used in the reasoner of KnowTD to find a way through this graph that connects the given variables and the target values that are to be calculated~\cite{KnowTD}. This approach limits the scope of problems that can currently be solved. The results of the advanced benchmark set show that some of the LLMs are still able to handle this kind of complex calculations. Therefore,  LLMs might be able to support KnowTD in this step of the reasoning process.

\paragraph{Using KnowTD to enhance the performance of LLMs}

There are two primary approaches for using knowledge-based systems to enhance the performance of LLMs: (i) They could be used in the training or fine-tuning of the LLM, or (ii) they could be used for advanced prompt
engineering and for the verification of answers of a given LLM.

In the first approach (i), the ontology of KnowTD could be used for training or fine-tuning an LLM. Recent research demonstrates that the vocabulary, relationships, and reasoning strategies embedded in ontologies can be integrated into LLMs to enhance their performance~\cite{FineTuningLLMsOntologyReasoning,ronzano2024, mai2024llmsreallyadaptdomains}. However, further investigation is needed to understand how ontology-based fine-tuning can improve the analytical reasoning required for complex problem-solving tasks. Also, knowledge-based problem solvers such as KnowTD could be exploited to generate a vast amount of sample exercises and verified solutions, which can be provided as training data for the LLMs. However, while such measures may enhance their performance, they do not fundamentally alter the inherently statistical nature of these models.

Following the second approach (ii), knowledge-based systems could enhance language models by helping to identify, correct, or even prevent the errors discussed in Section~\ref{Error Types}. 
Errors such as inconsistent signs or the use of incorrect equations can be detected by cross-referencing the language model's output with the validated equations stored in the KnowTD ontology. 
Verification could be performed at varying levels. At a basic level, the ontology can be used to confirm whether or not the equations applied by the LLM exist within the knowledge base. At a more advanced level, the ontology can also be queried to check the context of the equations and whether they are allowed to be used in the scenario at hand. If problems are identified, the LLM could be prompted to revise its response, to specify where the error occurred and eventually to also specify additional information on the type and context of the error. In prompt engineering, KnowTD could, for instance, be used to provide background information and applicable equations.

\section{Conclusion}
\label{conclusion}
In this work, we have investigated the capabilities of current LLMs for solving thermodynamic problems. Two sets of benchmark problems were used, a simple and an advanced one. The benchmark problems were designed for the present study and are fully disclosed, as well as detailed information on the assessment of two of the answers of the LLMs, which was carried out by trained experts in the same way as the answers of students would have been evaluated. The variability of the answers of the LLMs -- evaluated by running each problem three independent times -- was found to be considerable. The standard deviation of the scores obtained in the repetitions was of the order of 1/3 to 1/2 of the mean score, underlining the statistical nature of the responses of the LLMs. The errors that occurred are of different nature, ranging from insufficient understanding of the problem over wrong assumptions to numerical problems. Not only the number of errors but also their nature varied strongly, depending on the type of problem (simple / advanced) and on the model. As expected, there was a clear deterioration going from the simple to the advanced problems in all cases. Interestingly, among the LLMs that were applied to the advanced problems (GPT-4, GPT-4o, Llama 3.1, le Chat), the differences between the scores were reduced compared to the simple problems. Not a single LLM clearly stands out, as measured by the scores: In all cases, the differences between the mean scores of the models were smaller than the standard deviation of the repetitions. Some of the observed differences in performance may result from random variability and not necessarily indicate differences in capability. However, a clear improvement was observed going from GPT-3.5, which was tested only on the simple problems, to the more recent versions of GPT models. The development of LLMs is rapid, new versions can perform very differently from older ones. \par

The LLMs rarely completely failed, even on the advanced problems, and for some of the simple problems, some of the LLMs consistently gave correct answers, which is remarkable. These results suggest that thermodynamic knowledge was included in the training data of all the studied LLMs. However, overall, the results highlight a critical limitation: The lack of analytical skills significantly restricts the use of LLMs for solving complex problems in science and engineering, a verification of their results is always necessary. The benchmark problems we have provided enable tracking future developments of LLMs and also devising and evaluating advanced prompting strategies. We question whether relying on general-purpose LLMs alone will be sufficient for reliably solving problems from science and engineering. The option of providing the LLMs with additional input, e.g., text from teaching materials, and evaluating their performance on the same benchmark problems is interesting to explore. Another solution could be to combine the LLMs with knowledge-based systems to take advantage of the strengths of both approaches. We have recently developed a prototype of such a knowledge-based system for thermodynamics \cite{KnowTD} and discuss options for the hybridization of that system with LLMs. We view LLMs and knowledge-based systems as complementary partners rather than competitors.

\section*{Acknowledgements}
We gratefully acknowledge funding by Deutsche Forschungsgemeinschaft DFG in the frame of the Priority Program 2331 ``Machine Learning in Chemical Engineering''. We also thank Lars Kobel for his support in carrying out the problem solving work.



 \bibliographystyle{elsarticle-num}
 \sloppy
 \bibliography{cas-refs}

\begin{thebibliography}{10}
\expandafter\ifx\csname url\endcsname\relax
  \def\url#1{\texttt{#1}}\fi
\expandafter\ifx\csname urlprefix\endcsname\relax\def\urlprefix{URL }\fi
\expandafter\ifx\csname href\endcsname\relax
  \def\href#1#2{#2} \def\path#1{#1}\fi

\bibitem{PhysSys}
R.~Borst, H.~Akkermans, A.~Pos, J.~Top, The physsys ontology for physical systems, Proceedings Workshop Qualitative Reasoning '95, Amsterdam, NL: 11-21 (01 1995).

\bibitem{OntoCape}
J.~Morbach, A.~Yang, W.~Marquardt, \href{https://www.sciencedirect.com/science/article/pii/S0952197606001199}{Ontocape—a large-scale ontology for chemical process engineering}, Engineering Applications of Artificial Intelligence 20~(2) (2007) 147--161, special Issue on Applications of Artificial Intelligence in Process Systems Engineering.
\newblock \href {https://doi.org/https://doi.org/10.1016/j.engappai.2006.06.010} {\path{doi:https://doi.org/10.1016/j.engappai.2006.06.010}}.
\newline\urlprefix\url{https://www.sciencedirect.com/science/article/pii/S0952197606001199}

\bibitem{OntoKin}
F.~Farazi, J.~Akroyd, S.~Mosbach, P.~Buerger, D.~Nurkowski, M.~Salamanca, M.~Kraft, \href{https://doi.org/10.1021/acs.jcim.9b00960}{Ontokin: An ontology for chemical kinetic reaction mechanisms}, Journal of Chemical Information and Modeling 60~(1) (2020) 108--120, pMID: 31846323.
\newblock \href {http://arxiv.org/abs/https://doi.org/10.1021/acs.jcim.9b00960} {\path{arXiv:https://doi.org/10.1021/acs.jcim.9b00960}}, \href {https://doi.org/10.1021/acs.jcim.9b00960} {\path{doi:10.1021/acs.jcim.9b00960}}.
\newline\urlprefix\url{https://doi.org/10.1021/acs.jcim.9b00960}

\bibitem{propnet}
D.~Mrdjenovich, M.~K. Horton, J.~H. Montoya, C.~M. Legaspi, S.~Dwaraknath, V.~Tshitoyan, A.~Jain, K.~A. Persson, \href{https://www.sciencedirect.com/science/article/pii/S2590238519303881}{propnet: A knowledge graph for materials science}, Matter 2~(2) (2020) 464--480.
\newblock \href {https://doi.org/https://doi.org/10.1016/j.matt.2019.11.013} {\path{doi:https://doi.org/10.1016/j.matt.2019.11.013}}.
\newline\urlprefix\url{https://www.sciencedirect.com/science/article/pii/S2590238519303881}

\bibitem{OSMO}
M.~T. Horsch, C.~Niethammer, G.~Boccardo, P.~Carbone, S.~Chiacchiera, M.~Chiricotto, J.~D. Elliott, V.~Lobaskin, P.~Neumann, P.~Schiffels, M.~A. Seaton, I.~T. Todorov, J.~Vrabec, W.~L. Cavalcanti, \href{https://doi.org/10.1021/acs.jced.9b00739}{Semantic interoperability and characterization of data provenance in computational molecular engineering}, Journal of Chemical \& Engineering Data 65~(3) (2020) 1313--1329.
\newblock \href {http://arxiv.org/abs/https://doi.org/10.1021/acs.jced.9b00739} {\path{arXiv:https://doi.org/10.1021/acs.jced.9b00739}}, \href {https://doi.org/10.1021/acs.jced.9b00739} {\path{doi:10.1021/acs.jced.9b00739}}.
\newline\urlprefix\url{https://doi.org/10.1021/acs.jced.9b00739}

\bibitem{PSO}
H.~Cheong, A.~Butscher, \href{https://doi.org/10.1080/09544828.2019.1644301}{Physics-based simulation ontology: an ontology to support modelling and reuse of data for physics-based simulation}, Journal of Engineering Design 30~(10-12) (2019) 655--687.
\newblock \href {http://arxiv.org/abs/https://doi.org/10.1080/09544828.2019.1644301} {\path{arXiv:https://doi.org/10.1080/09544828.2019.1644301}}, \href {https://doi.org/10.1080/09544828.2019.1644301} {\path{doi:10.1080/09544828.2019.1644301}}.
\newline\urlprefix\url{https://doi.org/10.1080/09544828.2019.1644301}

\bibitem{QAChemistry}
X.~Zhou, D.~Nurkowski, S.~Mosbach, J.~Akroyd, M.~Kraft, \href{https://doi.org/10.1021/acs.jcim.1c00275}{Question answering system for chemistry}, Journal of Chemical Information and Modeling 61~(8) (2021) 3868--3880, pMID: 34338504.
\newblock \href {http://arxiv.org/abs/https://doi.org/10.1021/acs.jcim.1c00275} {\path{arXiv:https://doi.org/10.1021/acs.jcim.1c00275}}, \href {https://doi.org/10.1021/acs.jcim.1c00275} {\path{doi:10.1021/acs.jcim.1c00275}}.
\newline\urlprefix\url{https://doi.org/10.1021/acs.jcim.1c00275}

\bibitem{DigitalTwin}
J.~Akroyd, S.~Mosbach, A.~Bhave, M.~Kraft, Universal digital twin - a dynamic knowledge graph, Data-Centric Engineering 2 (2021) e14.
\newblock \href {https://doi.org/10.1017/dce.2021.10} {\path{doi:10.1017/dce.2021.10}}.

\bibitem{KnowTD}
L.~Vollmer, S.~Fellenz, F.~Jirasek, H.~Leitte, H.~Hasse, \href{https://doi.org/10.1021/acs.jcim.4c00647}{Knowtd--an actionable knowledge representation system for thermodynamics}, Journal of Chemical Information and Modeling 64~(15) (2024) 5878--5887, pMID: 39042488.
\newblock \href {http://arxiv.org/abs/https://doi.org/10.1021/acs.jcim.4c00647} {\path{arXiv:https://doi.org/10.1021/acs.jcim.4c00647}}, \href {https://doi.org/10.1021/acs.jcim.4c00647} {\path{doi:10.1021/acs.jcim.4c00647}}.
\newline\urlprefix\url{https://doi.org/10.1021/acs.jcim.4c00647}

\bibitem{Mathematica}
W.~R. Inc., \href{https://www.wolfram.com/mathematica}{Mathematica, {V}ersion 14.1}, champaign, IL, 2024.
\newline\urlprefix\url{https://www.wolfram.com/mathematica}

\bibitem{Yao_2024}
Y.~Yao, J.~Duan, K.~Xu, Y.~Cai, Z.~Sun, Y.~Zhang, A survey on large language model (llm) security and privacy: The good, the bad, and the ugly, High-Confidence Computing 4~(2) (2024) 100211.
\newblock \href {https://doi.org/10.1016/j.hcc.2024.100211} {\path{doi:10.1016/j.hcc.2024.100211}}.

\bibitem{vaswani2017attention}
A.~Vaswani, Attention is all you need, Advances in Neural Information Processing Systems (2017).

\bibitem{openai2024gpt}
K.~OpenAI, et~al., Gpt-4o system card, Tech. rep., OpenAI (2024).

\bibitem{llama_meta}
A.~.~M. Llama~Team, The llama 3 herd of models, Tech. rep., Meta (2024).

\bibitem{MistralAI_Website}
Mistral large 2, \url{https://mistral.ai/news/mistral-large-2407/}, accessed: 2024-10-02.

\bibitem{Tsai_2023}
M.-L. Tsai, C.~W. Ong, C.-L. Chen, Exploring the use of large language models (llms) in chemical engineering education: Building core course problem models with chat-gpt, Education for Chemical Engineers 44 (2023) 71--95.
\newblock \href {https://doi.org/10.1016/j.ece.2023.05.001} {\path{doi:10.1016/j.ece.2023.05.001}}.

\bibitem{wardat2023chatgpt}
Y.~Wardat, M.~A. Tashtoush, R.~AlAli, A.~M. Jarrah, Chatgpt: A revolutionary tool for teaching and learning mathematics, Eurasia Journal of Mathematics, Science and Technology Education 19~(7) (2023) em2286.

\bibitem{liang_exploring_2023}
Y.~Liang, D.~Zou, H.~Xie, F.~L. Wang, \href{https://doi.org/10.1186/s40561-023-00273-7}{Exploring the potential of using {ChatGPT} in physics education}, Smart Learning Environments 10~(1) (2023) 52.
\newblock \href {https://doi.org/10.1186/s40561-023-00273-7} {\path{doi:10.1186/s40561-023-00273-7}}.
\newline\urlprefix\url{https://doi.org/10.1186/s40561-023-00273-7}

\bibitem{krupp2024unreflected}
L.~Krupp, S.~Steinert, M.~Kiefer-Emmanouilidis, K.~E. Avila, P.~Lukowicz, J.~Kuhn, S.~K{\"u}chemann, J.~Karolus, Unreflected acceptance--investigating the negative consequences of chatgpt-assisted problem solving in physics education, in: HHAI 2024: Hybrid Human AI Systems for the Social Good, IOS Press, 2024, pp. 199--212.

\bibitem{Wang_2024_pitfallsOfChatGPT}
K.~D. Wang, E.~Burkholder, C.~Wieman, S.~Salehi, N.~Haber, Examining the potential and pitfalls of chatgpt in science and engineering problem-solving, Frontiers in Education 8 (2024).
\newblock \href {https://doi.org/10.3389/feduc.2023.1330486} {\path{doi:10.3389/feduc.2023.1330486}}.

\bibitem{Kortemeyer_2024_GradingAssistance}
G.~Kortemeyer, J.~N{\"o}hl, D.~Onishchuk, Grading assistance for a handwritten thermodynamics exam using artificial intelligence: An exploratory study, Physical Review Physics Education Research 20~(2) (2024).
\newblock \href {https://doi.org/10.1103/PhysRevPhysEducRes.20.020144} {\path{doi:10.1103/PhysRevPhysEducRes.20.020144}}.

\bibitem{Polverini_2024_UnderstandingGraphsKinematics}
G.~Polverini, B.~Gregorcic, Performance of chatgpt on the test of understanding graphs in kinematics, Physical Review Physics Education Research 20~(1) (2024).
\newblock \href {https://doi.org/10.1103/PhysRevPhysEducRes.20.010109} {\path{doi:10.1103/PhysRevPhysEducRes.20.010109}}.

\bibitem{Yeadon_2024_GCSE_universityLevels}
W.~Yeadon, T.~Hardy, The impact of ai in physics education: a comprehensive review from gcse to university levels, Physics Education 59~(2) (2024) 025010.
\newblock \href {https://doi.org/10.1088/1361-6552/ad1fa2} {\path{doi:10.1088/1361-6552/ad1fa2}}.

\bibitem{Sperling_2024_AI_HighScoolPhysics}
A.~Sperling, J.~Lincoln, Artificial intelligence and high school physics, The Physics Teacher 62~(4) (2024) 314--315.
\newblock \href {https://doi.org/10.1119/5.0202994} {\path{doi:10.1119/5.0202994}}.

\bibitem{plevris2023chatbots}
V.~Plevris, G.~Papazafeiropoulos, A.~Jim{\'e}nez~Rios, Chatbots put to the test in math and logic problems: a comparison and assessment of chatgpt-3.5, chatgpt-4, and google bard, AI 4~(4) (2023) 949--969.

\bibitem{frieder2024Mathematical}
S.~Frieder, L.~Pinchetti, A.~Chevalier, R.-R. Griffiths, T.~Salvatori, T.~Lukasiewicz, P.~Petersen, J.~Berner, Mathematical capabilities of chatgpt, in: Proceedings of the 37th International Conference on Neural Information Processing Systems, NIPS '23, Curran Associates Inc., Red Hook, NY, USA, 2024.

\bibitem{zhang_gap_2020}
D.~Zhang, L.~Wang, L.~Zhang, B.~T. Dai, H.~T. Shen, \href{https://ieeexplore.ieee.org/document/8703135/}{The {Gap} of {Semantic} {Parsing}: {A} {Survey} on {Automatic} {Math} {Word} {Problem} {Solvers}}, IEEE Transactions on Pattern Analysis and Machine Intelligence 42~(9) (2020) 2287--2305, number: 9.
\newblock \href {https://doi.org/10.1109/TPAMI.2019.2914054} {\path{doi:10.1109/TPAMI.2019.2914054}}.
\newline\urlprefix\url{https://ieeexplore.ieee.org/document/8703135/}

\bibitem{patel-etal-2021-nlp}
A.~Patel, S.~Bhattamishra, N.~Goyal, \href{https://aclanthology.org/2021.naacl-main.168}{Are {NLP} models really able to solve simple math word problems?}, in: K.~Toutanova, A.~Rumshisky, L.~Zettlemoyer, D.~Hakkani-Tur, I.~Beltagy, S.~Bethard, R.~Cotterell, T.~Chakraborty, Y.~Zhou (Eds.), Proceedings of the 2021 Conference of the North American Chapter of the Association for Computational Linguistics: Human Language Technologies, Association for Computational Linguistics, Online, 2021, pp. 2080--2094.
\newblock \href {https://doi.org/10.18653/v1/2021.naacl-main.168} {\path{doi:10.18653/v1/2021.naacl-main.168}}.
\newline\urlprefix\url{https://aclanthology.org/2021.naacl-main.168}

\bibitem{lu-etal-2023-survey}
P.~Lu, L.~Qiu, W.~Yu, S.~Welleck, K.-W. Chang, \href{https://aclanthology.org/2023.acl-long.817}{A survey of deep learning for mathematical reasoning}, in: A.~Rogers, J.~Boyd-Graber, N.~Okazaki (Eds.), Proceedings of the 61st Annual Meeting of the Association for Computational Linguistics (Volume 1: Long Papers), Association for Computational Linguistics, Toronto, Canada, 2023, pp. 14605--14631.
\newblock \href {https://doi.org/10.18653/v1/2023.acl-long.817} {\path{doi:10.18653/v1/2023.acl-long.817}}.
\newline\urlprefix\url{https://aclanthology.org/2023.acl-long.817}

\bibitem{venkatasubramanian2024whatsembeddingroseembedding}
V.~Venkatasubramanian, \href{https://arxiv.org/abs/2406.06870}{What's in an embedding? would a rose by any embedding smell as sweet?} (2024).
\newblock \href {http://arxiv.org/abs/2406.06870} {\path{arXiv:2406.06870}}.
\newline\urlprefix\url{https://arxiv.org/abs/2406.06870}

\bibitem{yao2024tree}
S.~Yao, D.~Yu, J.~Zhao, I.~Shafran, T.~Griffiths, Y.~Cao, K.~Narasimhan, Tree of thoughts: Deliberate problem solving with large language models, Advances in Neural Information Processing Systems 36 (2024).

\bibitem{Zhao_2024_LLM_Blackbox}
H.~Zhao, F.~Yang, B.~Shen, H.~Lakkaraju, M.~Du, Towards uncovering how large language model works: An explainability perspective.
\newblock \href {https://doi.org/10.48550/arXiv.2402.10688} {\path{doi:10.48550/arXiv.2402.10688}}.

\bibitem{Pursnani_2023_engineeringExamUSFundamentals}
V.~Pursnani, Y.~Sermet, M.~Kurt, I.~Demir, Performance of chatgpt on the us fundamentals of engineering exam: Comprehensive assessment of proficiency and potential implications for professional environmental engineering practice, Computers and Education: Artificial Intelligence 5 (2023) 100183.
\newblock \href {https://doi.org/10.1016/j.caeai.2023.100183} {\path{doi:10.1016/j.caeai.2023.100183}}.

\bibitem{Wang_2023_SciBench}
X.~Wang, Z.~Hu, P.~Lu, Y.~Zhu, J.~Zhang, S.~Subramaniam, A.~R. Loomba, S.~Zhang, Y.~Sun, W.~Wang, Scibench: Evaluating college-level scientific problem-solving abilities of large language models.
\newblock \href {https://doi.org/10.48550/arXiv.2307.10635} {\path{doi:10.48550/arXiv.2307.10635}}.

\bibitem{Song_01.06.2024}
T.~Song, Y.~Fan, C.~Feng, K.~Song, C.~Liu, D.~Jiang, \href{http://arxiv.org/pdf/2406.07572v1}{Domain-specific react for physics-integrated iterative modeling: A case study of llm agents for gas path analysis of gas turbines}.
\newline\urlprefix\url{http://arxiv.org/pdf/2406.07572v1}

\bibitem{Li2024_simulatingexperts}
Z.~Li, Y.~Chang, X.~Le, Simulating expert discussions with multi-agent for enhanced scientific problem solving, in: Proceedings of the Fourth Workshop on Scholarly Document Processing (SDP 2024), 2024, pp. 243--256.

\bibitem{OpenAI_Website_3.5}
Models: Gpt-3.5 turbo, \url{https://platform.openai.com/docs/models/gpt-3-5-turbo}, accessed: 2024-12-04.

\bibitem{OpenAI_Website_4}
Models: Gpt-4 turbo and gpt-4, \url{https://platform.openai.com/docs/models/gpt-4-turbo-and-gpt-4}, accessed: 2024-12-04.

\bibitem{OpenAI_Website_4o}
Models: Gpt-4o, \url{https://platform.openai.com/docs/models/gpt-4o}, accessed: 2024-12-04.

\bibitem{vavekanand2024llama}
R.~Vavekanand, K.~Sam, Llama 3.1: An in-depth analysis of the next-generation large language model (2024).

\bibitem{MistralAI_Dataset}
Does mistral ai communicate on the training datasets?, \url{https://help.mistral.ai/en/articles/156195-does-mistral-ai-communicate-on-the-training-datasets}, accessed: 2024-10-02.

\bibitem{xu2024surveymultilinguallargelanguage}
Y.~Xu, L.~Hu, J.~Zhao, Z.~Qiu, Y.~Ye, H.~Gu, \href{https://arxiv.org/abs/2404.00929}{A survey on multilingual large language models: Corpora, alignment, and bias} (2024).
\newblock \href {http://arxiv.org/abs/2404.00929} {\path{arXiv:2404.00929}}.
\newline\urlprefix\url{https://arxiv.org/abs/2404.00929}

\bibitem{Rudolph_2023}
J.~Rudolph, S.~Tan, S.~Tan, Chatgpt: Bullshit spewer or the end of traditional assessments in higher education?, Journal of Applied Learning {\&} Teaching 6~(1) (2023).
\newblock \href {https://doi.org/10.37074/jalt.2023.6.1.9} {\path{doi:10.37074/jalt.2023.6.1.9}}.

\bibitem{Haque_2024}
M.~A. Haque, S.~Li, Exploring chatgpt and its impact on society, AI and Ethics (2024).
\newblock \href {https://doi.org/10.1007/s43681-024-00435-4} {\path{doi:10.1007/s43681-024-00435-4}}.

\bibitem{Zaitsu_2023}
W.~Zaitsu, M.~Jin, Distinguishing chatgpt(-3.5, -4)-generated and human-written papers through japanese stylometric analysis, PloS one 18~(8) (2023) e0288453.
\newblock \href {https://doi.org/10.1371/journal.pone.0288453} {\path{doi:10.1371/journal.pone.0288453}}.

\bibitem{Shahriar_2024}
S.~Shahriar, B.~D. Lund, N.~R. Mannuru, M.~A. Arshad, K.~Hayawi, R.~V.~K. Bevara, A.~Mannuru, L.~Batool, Putting gpt-4o to the sword: A comprehensive evaluation of language, vision, speech, and multimodal proficiency, Applied Sciences 14~(17) (2024) 7782.
\newblock \href {https://doi.org/10.3390/app14177782} {\path{doi:10.3390/app14177782}}.

\bibitem{OpenAI_Website_o1-preview}
Introducing openai o1-preview, \url{https://openai.com/index/introducing-openai-o1-preview/}, accessed: 2024-10-18.

\bibitem{Lingo_14.09.2024}
R.~Lingo, M.~Arroyo, R.~Chhajer, \href{http://arxiv.org/pdf/2409.09415v1}{Enhancing llm problem solving with reap: Reflection, explicit problem deconstruction, and advanced prompting}.
\newline\urlprefix\url{http://arxiv.org/pdf/2409.09415v1}

\bibitem{de2024system}
J.~de~Winter, D.~Dodou, Y.~B. Eisma, System 2 thinking in openai's o1-preview model: Near-perfect performance on a mathematics exam, arXiv preprint arXiv:2410.07114 (2024).

\bibitem{FineTuningLLMsOntologyReasoning}
T.~Baldazzi, L.~Bellomarini, S.~Ceri, A.~Colombo, A.~Gentili, E.~Sallinger, Fine-tuning large enterprise language models via ontological reasoning, in: A.~Fensel, A.~Ozaki, D.~Roman, A.~Soylu (Eds.), Rules and Reasoning, Springer Nature Switzerland, Cham, 2023, pp. 86--94.

\bibitem{ronzano2024}
F.~Ronzano, J.~Nanavati, \href{https://arxiv.org/abs/2405.20527}{Towards ontology-enhanced representation learning for large language models} (2024).
\newblock \href {http://arxiv.org/abs/2405.20527} {\path{arXiv:2405.20527}}.
\newline\urlprefix\url{https://arxiv.org/abs/2405.20527}

\bibitem{mai2024llmsreallyadaptdomains}
H.~T. Mai, C.~X. Chu, H.~Paulheim, \href{https://arxiv.org/abs/2407.19998}{Do llms really adapt to domains? an ontology learning perspective} (2024).
\newblock \href {http://arxiv.org/abs/2407.19998} {\path{arXiv:2407.19998}}.
\newline\urlprefix\url{https://arxiv.org/abs/2407.19998}

\end{thebibliography}

\end{document}



\author[1]{Rébecca Loubet}
\author[1]{Pascal Zittlau}
\author[2]{Luisa Vollmer}
\author[1]{Marco Hoffmann}
\author[3]{Sophie Fellenz}
\author[1]{Fabian Jirasek}
\author[2]{Heike Leitte}
\author[1]{Hans Hasse\corref{cor1}}
\ead{hans.hasse@rptu.de}
\cortext[cor1]{Corresponding author}


\affiliation[1]{organization={Laboratory of Engineering Thermodynamics},
            addressline={RPTU Kaiserslautern}, 
            city={Kaiserslautern},
            postcode={67655}, 
            state={RP},
            country={Germany}}
            
\affiliation[2]{organization={Laboratory of Visual Information Analysis},
            addressline={RPTU Kaiserslautern}, 
            city={Kaiserslautern},
            postcode={67655}, 
            state={RP},
            country={Germany}}

\affiliation[3]{organization={Laboratory of Machine Learning},
            addressline={RPTU Kaiserslautern}, 
            city={Kaiserslautern},
            postcode={67655}, 
            state={RP},
            country={Germany}}


\maketitle  

\tableofcontents  

\section{Simple test cases}
	Below, we present the collection of simple test cases developed and used in this study, complete with numerical solutions and the number of points obtainable for each question.
	
	\subsection{Problem 1}
	
	A gas in a cylinder is compressed reversibly from v\_1~=~0.05~m\^{}3/kg to v\_2~=~0.02~m\^{}3/kg. The initial temperature is T\_1~=~298~K. The process is adiabatic. What is the work supplied per kilogram of gas?
	
	The gas is ideal, with R~=~287~J/(kg~K) and c\_v~=~1010~J/(kg~K).\\
	
	\textit{Solution:} w\_12~=~8.95$\times$10\^{}4~J/kg (3.5 points)
	
	\subsection{Problem 2}
	
	A gas in a cylinder undergoes a polytropic expansion with the exponent n~=~0.8. The initial pressure and specific volume are p\_1~=~2 bar and v\_1~=~0.7~m\^{}3/kg, the final specific volume is v\_2~=~1.1~m\^{}3/kg. Calculate the following quantities per kilogram of gas: work, heat, entropy change.
	
	The gas is ideal, with R~=~277~J/(kg~K) and c\_v~=~1746~J/(kg~K).\\
	
	\textit{Solution:} w\_12~=~-6.62$\times$10\^{}4~J/kg; q\_12~=~1.5$\times$10\^{}5~J/kg; delta\_s\_12~=~283.03~J/(kg~K) (5.5 points)
	
	\subsection{Problem 3}
	
	A gas in a cooled cylinder undergoes an isothermal compression at T~=~298~K. The initial and final pressure are p\_1~=~1~bar, p\_2~=~1.6~bar. Calculate the heat that has to be removed per kilogram of gas.
	
	The gas is ideal, with R~=~518~J/(kg~K).\\
	
	\textit{Solution:} q\_12~=~-7.26$\times$10\^{}4~J/kg (2.5 points)
	
	\subsection{Problem 4}
	
	A gas in a heated cylinder is expanded. The pressure changes from p\_1~=~1.7 bar to p\_2~=~1.05~bar, the specific volume changes from and v\_1~=~0.2~m\^{}3/kg to v\_2~=~0.5~m\^{}3/kg, the work obtained per kilogram of gas is 8.7~kJ/kg. Calculate the heat that is supplied per kilogram of gas.
	
	The gas is ideal, with R~=~117~J/(kg~K) and c\_v~= 479~J/(kg~K).\\
	
	\textit{Solution:} q\_12~=~8.44$\times$10\^{}4~J/kg  (3 points)
	
	\subsection{Problem 5}
	
	A gas in a cylinder is expanded, whereby the specific volume increases from v\_1~=~0.8 m\^{}3/kg to v\_2~=~1.5~m\^{}3/kg. The cylinder is heated in such a way that the pressure p = 1 bar remains constant during the entire process. Calculate the work and the heat per kilogram of gas.
	
	The gas is ideal, with R~=~297~J/(kg~K) and c\_v~=~1040~J/(kg~K).\\
	
	\textit{Solution:} w\_12~=~-7$\times$10\^{}4~J/kg; q\_12~=~3.15$\times$10\^{}5~J/kg (4 points)
	
	\subsection{Problem 6}
	
	A gas in rigid tank with the volume V~=~0.05~m\^{}3 is stirred. The initial temperature and pressure are T\_1~=~315~K, p\_1~=~1~bar. The stirrer supplies a work of 0.5~kJ. The tank is cooled, as there is a threshold for the final temperature. Calculate the heat that is removed if the final temperature is T\_2~=~320 K.
	
	The gas is ideal, with R~=~287~J/(kg~K) and c\_v~=~1010~J/(kg~K).\\
	
	\textit{Solution:} Q\_12~=~-2.21$\times$10\^{}2~J/kg (2.5 points)
	
	\subsection{Problem 7}
	
	A gas in a cylinder is compressed isentropically. The pressure and specific volume change from p\_1~=~1 bar, v\_1~=~0.52~m\^{}3/kg to p\_2~=~3~bar, v\_2~=~0.2~m\^{}3/kg. Calculate the molar heat capacity at constant volume c\_v of the gas under the assumption that c\_v is constant.
	
	The gas is ideal, its molar mass is M~=~44 g/mol.\\
	
	\textit{Solution:} c\_v~=~55.51~J/(mol~K) (3.5 points)
	
	\subsection{Problem 8}
	
	A cylinder contains m~=~1 kg of an ideal gas. The initial volume, pressure and temperature are V\_1~=~0.8~m\^{}3, p\_1~=~1~bar, and T\_1~=~298~K. After an adiabatic reversible compression, the final pressure and temperature are p\_2~=~2.9~bar, and T\_2~=~333~K. Calculate the molar mass of the gas, and its molar heat capacity at constant volume c\_v.\\
	
	\textit{Solution:} M~=~0.031~kg/mol; c\_v~=~71.54~J/(mol~K) (5 points)
	
	\subsection{Problem 9}
	
	A gas in a cylinder is compressed adiabatically. The initial temperature and specific volume are T\_1~=~298~K and v\_1~=~0.8~m\^{}3/kg. The final volume is v\_2~=~0.5~m\^{}3/kg. Calculate the minimal work that is required per kilogram of gas for this compression.
	
	The gas is ideal, with R~=~297~J/(kg~K) and c\_v~=~1040~J/(kg~K).\\
	
	\textit{Solution:} w\_12~=~4.45$\times$10\^{}4~J/kg (3.5 points)
	
	\subsection{Problem 10}
	
	A gas with the density rho~=~1.2~kg/m\^{}3 in a rigid adiabatic tank with the volume V~=~0.01~m\^{}3 is stirred. The initial temperature is T\_1~=~298~K. Calculate the temperature T\_2 in the tank, after the stirrer has supplied a work of 1~kJ.
	
	The gas is ideal with c\_v~=~1746~J/(kg~K).\\
	
	\textit{Solution:} T\_2~=~345.73~K (3 points)
	
	\subsection{Problem 11}
	
	A gas in a closed rigid cylinder is heated by up from T\_1~=~298~K to T\_2~=~353~K. The supplied heat is 1.3~kJ. Calculate the mass of the gas in the cylinder.
	
	The gas is ideal with c\_v~=~479~J/(kg~K).\\
	
	\textit{Solution:} m~=~0.049~kg (2 points)
	
	\subsection{Problem 12}
	
	A gas in a cylinder expands adiabatically. The initial pressure and specific volume are p\_1~=~1.9 bar and v\_1~=~0.9~m\^{}3/kg, the final pressure and specific volume are p\_2~=~1~bar and v\_2~=~1.5~m\^{}3/kg. Calculate the work per kilogram of gas and the change of the specific entropy.
	
	The gas is ideal with c\_v~=~1040~J/(kg~K) and has a molar mass of 28~g/mol.\\
	
	\textit{Solution:} w\_12~=~-7.36$\times$10\^{}4~J/kg; delta\_s\_12~=~15.41~J/(kg~K) (4.5 points)
	
	\subsection{Problem 13}
	
	A gas in a cylinder undergoes an adiabatic polytropic expansion. The initial pressure and temperature are p\_1~=~2.1~bar and T\_1~=~273~K, the final pressure is p\_2~=~1~bar. Calculate the work per kilogram of gas.
	
	The gas is ideal with c\_v~=~1009~J/(kg~K) and has a molar mass of 30~g/mol.\\
	
	\textit{Solution:} w\_12~=~-4.07$\times$10\^{}4~J/kg (4 points)
	\newpage
	
	\section{Advanced test cases}
	In the following, we present the collection of advanced test cases developed and used in this study, complete with numerical solutions and the number of points obtainable for each question.
    
	\subsection{Problem 14}
	
	A balloon is filled with helium. In state 1, the pressure and temperature of the helium are p\_1 = 1 bar and T\_1~=~280~K and the volume is V\_1 = 0.6 m\^{}3. After cooling (state 2), the temperature is T\_2 = 270 K and the balloon reaches its maximum volume V\_2 = 0.8 m\^{}3. The helium in the balloon is then heated while the volume remains constant, the pressure after the heating (state 3) is p\_3 = 0.2 bar higher than in state 2. Starting from state 3, helium is released from the balloon, whereby the volume of the balloon remains constant. The pressure p\_4 in the state 4 after the release is equal to the pressure p\_2. The change of state of the helium that remains in the balloon during the release is adiabatic and reversible.\\
	\begin{enumerate}[label=\alph*)]
		\item Calculate the initial mass of helium in the balloon m\_1 and the pressure p\_2.
		\item Calculate the heat that is supplied to the helium during the isochoric heating.
		\item Calculate the temperature T\_4 and the mass of helium that is still in the balloon in state 4.
	\end{enumerate}\vspace*{10pt}
	\textit{Assumptions:}
	\begin{itemize}
		\item Changes in the kinetic and potential energy are negligible.
	\end{itemize}
	\textit{Fluid properties:}
	\begin{itemize}
		\item Helium is a perfect gas with R = 2.1 kJ/(kg K) and c\_v = 5.2 kJ/(kg K).
	\end{itemize}
	\textit{Solutions:}
	\begin{enumerate}[label=\alph*)]
		\item m\_1~=~1.02$\times$10\^{}-1~kg; p\_2~=~7.23$\times$10\^{}4~Pa (2 points)
		\item Q\_23~=~3.96$\times$10\^{}4~J (3 points)
		\item T\_4~=~321.29~K; m\_4~=~8.57$\times$10\^{}-2~kg (2.5 points)
	\end{enumerate}
	\newpage
	
	\subsection{Problem 15}
	
	A cylinder is divided into two compartments by a piston. Compartment A is filled with a liquid and compartment B with an ideal gas. Compartment B is connected to a second cylinder C by a pipe with a valve. The cylinder C is filled with the same gas as compartment B. The volume of cylinder C is 0.05 m\^{}3.\\[10pt]
	In the initial sate, the piston is locked and the pressures in both compartments is 1 bar, the
	valve between compartment B and cylinder C is closed, and the pressure in cylinder C is 8 bar. Then, the valve is opened; after the equilibration (state 2) the pressure in compartment B and cylinder C is 1.5 bar. The piston is then unlocked; in the final state 3 the pressure is 1.49 bar both in the liquid and in the gas.\\[10pt]
	The temperature is 293 K throughout.\\	
	\begin{enumerate}[label=\alph*)]
		\item Calculate for the initial state 1: the mass of the gas in cylinder C, the mass of the gas in compartment B and the volume of compartment B.
		\item Calculate the volume change and the heat supplied to the gas during the transition 2 $\rightarrow$ 3.
		\item Calculate the mass of the liquid in compartment A.
	\end{enumerate}\vspace*{10pt}
	
	\textit{Assumptions:}
	\begin{itemize}
		\item The piston moves frictionless.
		\item Changes in the kinetic and potential energy are negligible.
		\item The volume of the gas in the pipe and in the valve is negligible.
	\end{itemize}\vspace*{10pt}
	\textit{Fluid properties:}
	\begin{itemize}
		\item The gas is a perfect gas with R = 0.285 kJ/(kg K).
		\item The compressibility of the liquid is given by ($\partial$v/$\partial$p)T = -2.23 x 10\^{}-11 m\^{}3/(kg Pa).
	\end{itemize}
	\textit{Solutions:}
	\begin{enumerate}[label=\alph*)]
		\item m\_C\_1~=~4.79$\times$10\^{}-1~kg; m\_B\_1~=~7.78$\times$10\^{}-1~kg; V\_B\_1~=~6.5$\times$10\^{}-1~m\^{}3 (3.5 points)
		\item delta\_V\_BC\_23~=~4.7$\times$10\^{}-3~m\^{}3; Q\_BC\_23~=~7.47$\times$10\^{}2~J (4.5 points)
		\item m\_A~=~4576~kg (4 points)
	\end{enumerate}
	\newpage

	\subsection{Problem 16}
	A liquid is stored in a cylinder which is closed by a movable piston. In state 1, the piston is locked and the volume of the cylinder is V\_1 = 7 L. The temperature and pressure are T\_1 = 308 K and p\_1 = 100 bar. In a first step, the liquid is cooled isochorically to T\_2 = 298 K (state 2). The locking mechanism of the piston is then released and the liquid is isothermally compressed by the piston. The final pressure is again p\_3 = 100 bar (state 3).\\
	\begin{enumerate}[label=\alph*)]
		\item Calculate the heat that is removed during the change of state 1 $\rightarrow$ 2 and the pressure p\_2.
		\item Calculate the final volume V\_3 and the work supplied to the liquid during the change of state 2 $\rightarrow$ 3.
	\end{enumerate}\vspace*{10pt}
	
	\textit{Properties of the liquid:}
	\begin{itemize}
		\item Density (308 K, 100 bar): rho = 800 kg/m\^{}3
		\item Thermal expansion coefficient: alpha = 8*10\^{}-4 K\^{}-1
		\item Thermal pressure coefficient: beta = 0.25 K\^{}-1
		\item Specific isochoric heat capacity: c\_v = 2800 J/(kg K).
	\end{itemize}
	It is assumed that alpha, beta and c\_v are constant in the relevant region.\\
	
	\textit{Solutions:}
	\begin{enumerate}[label=\alph*)]
		\item Q\_12~=~-1.57$\times$10\^{}5~J; p\_2~=~8.21$\times$10\^{}5~Pa (5.5 points)
		\item V\_3~=~7.02$\times$10\^{}-3~m\^{}3; W\_i\_23~=~2.05$\times$10\^{}2~J (6.5 points)
	\end{enumerate}
	\newpage
	
	\subsection{Problem 17}
	The following open Joule process is considered here:\\[10pt]
	\begin{tabularx}{\textwidth}{lX}
		1 $\rightarrow$ 2:& Isentropic pressure increase, starting from p\_1 = 2 bar and T\_1 = 300 K to p\_2 = 10 bar. \\[10pt]
		2 $\rightarrow$ 3:& Isobaric heating; the heat supplied per kilogram of the working fluid is 50~kJ/kg. \\[10pt]
		3 $\rightarrow$ 4:& Isentropic pressure decrease to p\_4 = p\_1.
	\end{tabularx}\\[10pt]
	Two different working fluids are considered:\\[10pt]
	\begin{tabularx}{\textwidth}{lX}
		I: & Perfect gas\\
		II: & Constant density fluid
	\end{tabularx}\\[10pt]
	\begin{enumerate}[label=\alph*)]
		\item Calculate the work obtained from the process per kilogram of working fluid for both working fluids I and II.
		\item Calculate the thermal efficiency of the process for both working fluids I and II.
	\end{enumerate}\vspace*{10pt}
	\textit{Assumptions:}
	\begin{itemize}
		\item Changes in the kinetic and potential energy are negligible.
	\end{itemize}\vspace*{10pt}
	
	\textit{Fluid properties:}\\[10pt]
	Working fluid I:
	\begin{itemize}
		\item c\_p\_I = 1.015 kJ/(kg K)
		\item R\_I = 0.29 kJ/(kg K)
	\end{itemize}
	Working fluid II:
	\begin{itemize}
		\item c\_II = 4.2 kJ/(kg K)
		\item rho\_II = 997 kg/m\^{}3
	\end{itemize}
	\textit{Solutions:}
	\begin{enumerate}[label=\alph*)]
		\item w\_t\_out\_I~=~-0.18$\times$10\^{}5~J/kg;  w\_t\_out\_II~=~0~J/kg (9.5 points)
		\item eta\_I~=~0.37; eta\_II~=~0 (2 points)
	\end{enumerate}
	
	\newpage
	
	\subsection{Problem 18}
	In a test rig, two streams of water are taken from two tanks (1 and 2) in which the water is stored at ambient conditions (T\_1 = T\_2 = 300 K, p\_1 = p\_2 = 1 bar). Tank 1 is located at the height of the rig, tank 2 is at a height of z\_2 = 30 m above the rig. The mass flow rates of both streams are equal. The water from tank 1 is isobarically heated in a heat exchanger, the water leaving the heat exchanger in state 3 is fed into a first inlet of an adiabatic mixer. The water from the second tank passes a line leading it to the rig and is then fed into the second inlet of the mixer in state 4. The water that leaves the mixer in state 5 is then fed into an isentropic pump with a mechanical power of 260 W. The water leaving the pump in state 6 is accelerated in an isentropic nozzle. The velocity of the water leaving the nozzle in state 7 is c\_7 = 30 m/s, the pressure is p\_7 = 1 bar, and the temperature is T\_7 = 330~K. The cross-section of the outlet of the nozzle is A\_7 = 0.2 cm\^{}2.\\
	\begin{enumerate}[label=\alph*)]
		\item Calculate the mass flow rate of the stream leaving the nozzle.
		\item Calculate the heat flow in the heat exchanger and the resulting temperature T\_3.
		\item Calculate the pressures p\_5 and p\_6.
	\end{enumerate}\vspace*{10pt}
	
	\textit{Assumptions:}
	\begin{itemize}
		\item The kinetic energy in states 1 to 6 is negligible.
		\item All connecting lines are adiabatic.
		\item The process is stationary.
	\end{itemize}\vspace*{10pt}
	
	\textit{Fluid properties:}
	\begin{itemize}
		\item Water is considered to be a constant density fluid with rho = 1000 kg/m\^{}3 and c = 4.2 kJ/(kg K).
		\item The gravitational acceleration is g = 9.81 m/s\^{}2.
	\end{itemize}
	\textit{Solutions:}
	\begin{enumerate}[label=\alph*)]
		\item m\_7~=~0.6~kg/s (1 point)
		\item Q\_dot\_13~=~7.53$\times$10\^{}4~W; T\_3~=~359.73~K (6.5 points)
		\item p\_5~=~5.5$\times$10\^{}5~Pa; p\_6~=~1.17$\times$10\^{}5~Pa (4 points)
	\end{enumerate}
	\newpage
	
	\subsection{Problem 19}
	A gas flows through a pipe with a mass flow rate of 100 kg/s. To compensate for pressure losses, a compressor station is operated in which an adiabatic compressor and an isentropic diffuser are connected in series. The gas is fed into the compressor with T\_1 = 289 K and p\_1 = 3.2 bar (state 1). The pressure of the gas after the adiabatic compression is p\_2 = 4 bar (state 2), the outlet pressure of the diffuser is p\_3 = 4.2 bar and the outlet velocity is c\_3 = 25 m/s (state 3). The diameter of the pipe is d\_1 = d\_3 = 1.5 m.\\
	\begin{enumerate}[label=\alph*)]
		\item Calculate the velocity c\_1 of the gas at the inlet of the compressor station and the temperature T\_3 at the outlet.
		\item Calculate the drive power of the compressor and the change of the specific entropy delta\_s\_12 in the compressor.
	\end{enumerate}\vspace*{10pt}
	
	\textit{Assumptions:}
	\begin{itemize}
		\item Changes in the potential energy are negligible.
		\item The process is stationary.
	\end{itemize}\vspace*{10pt}
	
	\textit{Fluid properties:}
	\begin{itemize}
		\item The gas is a perfect gas with R = 0.518 kJ/(kg K) and c\_p = 2.24 kJ/(kg K).
	\end{itemize}
	\textit{Solutions:}
	\begin{enumerate}[label=\alph*)]
		\item c\_1~=~26.47~m/s; T\_3~=~358.21~K (3.5 points)
		\item W\_t\_dot\_12~=~1.55$\times$10\^{}7~W; delta\_s\_12~=~3.4$\times$10\^{}2~J/(kg~K) (5 points)
	\end{enumerate}
	\newpage
	
	\subsection{Problem 20}
	A heating circuit is operated with liquid water. Heat is supplied to the water from two sources: a solar collector and a flue gas-water heat exchanger (A), in which hot flue gas coming from a burner is cooled down by the water. The circuit also contains a throttle, a pump and a heat exchanger (B) in which the circulating water is cooled down and thereby supplies heat to consumers.\\[10pt]
	The mass flow rate of the water in the circuit is 0.15 kg/s.\\[10pt]
	In the circuit, the water undergoes the following changes of state:\\[10pt]
	\begin{tabularx}{\textwidth}{lX}
		1 $\rightarrow$ 2:& Isentropic pressure increase in the pump from p\_1 = 2 bar, T\_1 = 363 K to p\_2 = 8 bar \\
		2 $\rightarrow$ 3:& Isobaric heating in the heat exchanger A to T\_3 = 403 K\\
		3 $\rightarrow$ 4:& Isobaric cooling in the heat exchanger B to T\_4 = 333 K\\
		4 $\rightarrow$ 5:& Adiabatic expansion in the throttle to p\_5 = p\_1\\
		5 $\rightarrow$ 1:& Isobaric heating in the solar collector.
	\end{tabularx}\\[10pt]
	
	The flue gas side of the process is modeled as follows:\\[10pt]
	\begin{tabularx}{\textwidth}{lX}
		6 $\rightarrow$ 7:& Isobaric heating. The inlet temperature is T\_6 = 606 K, the supplied heat is 30 kW.\\
		7 $\rightarrow$ 8:& Isobaric cooling in the gas/water heat exchanger to T\_8 = 646 K.
	\end{tabularx}\\[10pt]
	\begin{enumerate}[label=\alph*)]
		\item Calculate the mechanical power of the pump.
		\item Calculate the heat flow in the heat exchanger B.
		\item Calculate the heat flow in the solar collector.
		\item Calculate the mass flow rate of the flue gas and its outlet temperature T\_7.
	\end{enumerate}\vspace*{10pt}
	
	\textit{Assumptions:}
	\begin{itemize}
		\item Changes in the kinetic and potential energy are negligible.
		\item The process is stationary.
	\end{itemize}\vspace*{10pt}
	
	\textit{Fluid properties:}
	\begin{itemize}
		\item Water is constant density fluid with rho\_W = 1000 kg/m\^{}3 and c\_W = 4.1 kJ/(kg K).
		\item The flue gas is a perfect gas with c\_p = 0.85 kJ/(kg K).
	\end{itemize}
	
	\textit{Solutions:}
	\begin{enumerate}[label=\alph*)]
		\item W\_t\_dot\_12~=90~W (2.5 points)
		\item Q\_dot\_34~=~-4.31$\times$10\^{}4~W (1 point)
		\item Q\_dot\_51~=~1.84$\times$10\^{}4~W (2 points)
		\item m\_G~=~1.59$\times$10\^{}-1~kg/s; T\_7~=~828.22~K (3.5 points)
	\end{enumerate}
	\newpage
	
	\subsection{Problem 21}
	In a remote party cabin, a heat engine is powered by biogas produced from kitchen waste. A generator converts 95\% of the mechanical power of the heat engine into electricity, which is used to operate an electric grill (power 3~kW) and a camping refrigerator. The waste heat flow from the heat engine is 3.1 kW and is used to heat the cabin. In the heat engine, a working fluid undergoes the following cyclic process in a cylinder closed by a piston:\\[10pt]
	
	\begin{tabularx}{\textwidth}{lX}
		1 $\rightarrow$ 2:& Adiabatic compression to T\_2 = 900 K.\\
		2 $\rightarrow$ 3:& Isobaric heating to T\_3 = 1800 K.\\
		3 $\rightarrow$ 4:& Adiabatic expansion to T\_4 = T\_2.\\
		4 $\rightarrow$ 1:& Isochoric cooling to T\_1, whereby a heat of q\_41 = 430.8 kJ per kilogram of working fluid is removed.
	\end{tabularx}\\[10pt]
	\begin{enumerate}[label=\alph*)]
		\item Calculate the net work per kilogram of working fluid obtained from the heat engine and its
		thermal efficiency.
		\item Calculate the electrical power supplied to the refrigerator.
		\item Calculate the largest possible coefficient of performance (COP) of a refrigerator that absorbs heat at 278~K and releases heat at 293 K. If such a refrigerator were available in the cabin, how much cooling power would it have and how much waste heat would it produce?
	\end{enumerate}\vspace*{10pt}
	
	\textit{Assumptions:}
	\begin{itemize}
		\item Changes in the kinetic and potential energy are negligible.
	\end{itemize}\vspace*{10pt}
	
	\textit{Fluid properties:}
	\begin{itemize}
		\item The working medium is a perfect gas with R = 288 J/(kg K) and c\_v = 718 J/(kg K).
	\end{itemize}
	
	\textit{Solutions:}
	\begin{enumerate}[label=\alph*)]
		\item w\_i\_netto~=~-4.75$\times$10\^{}5~J/kg; eta~=~0.524 (7.5 points)
		\item W\_t\_dot\_heat\_engine~=~-3.42$\times$10\^{}3~W (2 points)
		\item COP\_max~=~18.54; Q\_dot\_refrigirator\_in~=~4.53$\times$10\^{}3~W; Q\_dot\_refrigirator\_out~=~-4.77$\times$10\^{}3~W (3 points)
	\end{enumerate}
	\newpage
	
	\subsection{Problem 22}
	The working fluid in a refrigerator undergoes the following changes of state:\\[10pt]
	\begin{tabularx}{\textwidth}{lX}
		1 $\rightarrow$ 2:& Isobaric cooling to T\_3 = T\_1, the heat per kilogram released from the working fluid is 120 kJ/kg.\\
		3 $\rightarrow$ 4:& Further isobaric cooling to T\_4 in an internal heat exchanger.\\
		4 $\rightarrow$ 5:& Adiabatic reversible expansion in a turbine to p\_5 = p\_1.\\
		5 $\rightarrow$ 6:& Isobaric heating up to T\_6 = 243 K. The heat flow cools the cold store.\\
		6 $\rightarrow$ 1:& Isobaric heating in the internal heat exchanger. 
	\end{tabularx}\\[10pt]
	
	The mechanical power of the turbine is entirely used to drive the compressor. The required additional power for the compressor is supplied electrically.\\
	\begin{enumerate}[label=\alph*)]
		\item Calculate the work input to the compressor per kilogram of working fluid.
		\item Calculate the temperature T\_4 and the work obtained in the turbine per kilogram of working
		fluid.
		\item Calculate the coefficient of performance (COP) of the process.
	\end{enumerate}\vspace*{10pt}
	
	\textit{Assumptions:}
	\begin{itemize}
		\item Changes in the kinetic and potential energy are negligible.
		\item The process is stationary.
	\end{itemize}\vspace*{10pt}
	
	\textit{Fluid properties:}
	\begin{itemize}
		\item The working fluid is a perfect gas with R = 0.287 kJ/(kg K) and c\_p = 1.01 kJ/(kg K).
	\end{itemize}
	\textit{Solutions:}
	\begin{enumerate}[label=\alph*)]
	\item w\_t\_12~=~-1.2$\times$10\^{}5~J/kg (3.5 points)
	\item T\_4~=~243~K; w\_t\_45~=~-6.58$\times$10\^{}4~J/kg (4.5 points)
	\item COP~=~1.21 (2.5 points)
	\end{enumerate}	
	
	\newpage
	\section{Examples for the assessment of the answers of the LLMs}
	We report here two examples for how the answers of the different LLMs were assessed, one for a simple problem and one for an advanced problem. Correct assumptions, formulas, and calculations are displayed in green and awarded the specified number of points, whereas incorrect ones, along with their corrections, are shown in red.
	\subsection{Problem 1: Answer of GPT 4 in run 2}
	
	\columnratio{0.9} 
	\begin{paracol}{2}	
		
		To solve this problem, we can use the first law of thermodynamics for an adiabatic process with
		an ideal gas, and the relationship between temperature and volume in such a process. For an
		ideal gas undergoing a \point{reversible adiabatic} process, the following relation holds:
		\begin{center} p\_1 * v\_1\^{}gamma = p\_2 * v\_2\^{}gamma \end{center}
		where gamma is the adiabatic index, which is the ratio of the specific heat at constant
		pressure (c\_p) to the specific heat at constant volume (c\_v):
		\begin{center} gamma = c\_p / c\_v \end{center}
		Since the \point{gas is ideal}, we can use the relation between c\_p, c\_v and R:
		\begin{center} c\_p = c\_v + R \end{center}
		Now we can calculate gamma:
		\begin{center} gamma = (c\_v + R) / c\_v \end{center}
		Once we have gamma, we can relate the temperatures and volumes using the following relation for an
		adiabatic process:
		\begin{center} \point{T\_1 * v\_1\^{}(gamma - 1) = T\_2 * v\_2\^{}(gamma - 1)} \end{center}
		We can solve for T\_2. After obtaining T\_2, we can calculate the work done (W) on the gas using
		the relation for work in an adiabatic process:
		\begin{center} W = c\_v * (T\_2 - T\_1) \end{center}
		Let's go ahead and calculate gamma first.
		The adiabatic index gamma is approximately 1.284.
		
		Next, we'll use the adiabatic process relation between temperatures and volumes:
		\begin{center} T\_1 * v\_1\^{}(gamma - 1) = T\_2 * v\_2\^{}(gamma - 1) \end{center}
		Let's rearrange the equation and solve for T\_2:
		\begin{center} \point{T\_2 = T\_1 * (v\_1/v\_2)\^{}(gamma - 1)} \end{center}
		Plugging in the values T\_1 = 298 K, v\_1 = 0.05 m\^{}3/kg, and v\_2 = 0.02 m\^{}3/kg, we can find T\_2.
		The final temperature ( T\_2 ) after compression is approximately 386.63 K.
		Finally, we'll calculate the work done (W) on the gas using: 
		\begin{center} \point{W = c\_v * (T\_2 - T\_1)}  \end{center}	
		Let's compute the work (W).
		The work supplied per kilogram of gas during the adiabatic compression is approximately
		\point{	89515.15 J/kg}.
		
		\switchcolumn
		\vspace{2\baselineskip} 	
		\point{1 point}\\
		\vspace{2.35cm}\\
		\point{0.5 points}\\
		\vspace{3.1cm}\\
		\point{0.5 points}\\
		\vspace{4.3cm}\\
		\point{0.5 points}\\
		\vspace{1.35cm}\\
		\point{0.5 points}\\
		\vspace{0.2cm}\\
		\point{0.5 points}	
        \vspace{0.2cm} \point{Overall score: 3.5 / 3.5}
    \end{paracol}
    \newpage
	
	\subsection{Problem 18: Answer of GPT 4o in run 1}
	
	\columnratio{0.9} 
	\begin{paracol}{2}	
		
		Given the problem, we need to find the solutions step by step. We'll start by documenting the key formulas and known values required for each part of the problem.
		
		\vspace{10pt}Fluid properties and constants:
		
		rho = 1000 kg/m\^{}3
		
		c = 4.2 kJ/(kg K) = 4200 J/kg·K
		
		g = 9.81 m/s\^{}2
		
		\vspace{10pt}Known values:
		
		T\_1 = T\_2 = 300 K
		
		p\_1 = p\_2 = 1 bar = 100000 Pa
		
		z\_2 = 30 m
		
		Mechanical power of pump (W\_pump) = 260 W
		
		c\_7 = 30 m/s
		
		p\_7 = 1 bar = 100000 Pa
		
		T\_7 = 330 K
		
		A\_7 = 0.2 cm\^{}2 = 0.00002 m\^{}2
		
		\vspace{10pt}a) Calculate the mass flow rate of the stream leaving the nozzle.
		
		\vspace{10pt}The mass flow rate (m\_dot) leaving the nozzle can be calculated using the equation for mass flow rate through an area:
		\begin{center}
			\point{m\_dot = rho * A * c}
		\end{center}
		Given values for state 7:
		
		A\_7 = 0.00002 m\^{}2
		
		c\_7 = 30 m/s
		
		\vspace{10pt}So, the mass flow rate m\_dot\_7 is:
		\begin{center}
			m\_dot\_7 = 1000 kg/m\^{}3 * 0.00002 m\^{}2 * 30 m/s
		\end{center}
		\vspace{10pt}Let's calculate this:
		
		\vspace{10pt}The mass flow rate of the stream leaving the nozzle (m\_dot\_7) is \point{0.6 kg/s}.
		
		\vspace{10pt}b) Calculate the heat flow in the heat exchanger and the resulting temperature T\_3.
		
		\vspace{10pt}Given:
		
		- The mass flow rates of both streams are equal, so \fault{m\_dot\_1 = m\_dot\_2 = m\_dot\_7}.
		
		- The specific heat capacity c = 4200 J/(kg K)
		
		- Initial temperature T\_1 = 300 K
		
		- Final temperature T\_7 = 330 K for one stream
		
		- The process is isobaric and the heat exchanger is assumed to have 100 \% effectiveness.
		
		\vspace{10pt}The heat flow (Q) in the heat exchanger is given by:
		\begin{center}
			\point{Q = m\_dot * c * (T\_3 - T\_1)}
		\end{center}
		We need to find T\_3.
		
		\vspace{10pt}First, let’s assume
		\begin{center}
			\fault{m\_dot\_1 = m\_dot = m\_dot\_7 = 0.6 kg/s.}
		\end{center}
		
		The heat added to the stream from Tank 1 to reach state 7 will be the same amount extracted from the heat exchanger to elevate temperature from T\_1 to T\_7.
		\begin{center}
			\fault{Q = m\_dot\_7 * c * (T\_7 - T\_1)}
		\end{center}
		Let's calculate this:

		The heat flow in the heat exchanger (Q) is \fault{75600 J/s} (or 75.6 kW), and the resulting temperature T\_3 is \fault{330 K}.
		
		\vspace{10pt}c) Calculate the pressures p\_5 and p\_6.
		
		\vspace{10pt}To determine p\_5 and p\_6, we'll consider the energy balance and the mechanical power of the pump.
		
		\vspace{10pt}For the pump (state 5 to state 6), the mechanical power (W\_pump) is given by:
		\begin{center}
			\point{W\_pump = m\_dot * (h\_6 - h\_5)}
		\end{center}
		
		Here, h\_6 - h\_5 represents the change in specific enthalpy. Assuming an isentropic process, changes in enthalpy for an incompressible fluid (water) can be approximated as:
		\begin{center}
			\point{h\_6 - h\_5 = v * (p\_6 - p\_5)}
		\end{center}
		Where:
		
		- v is the specific volume of water, which is approximately the reciprocal of density (v = 1 / rho)
		
		- rho = 1000 kg/m\^{}3
		
		\vspace{10pt}So, we have:
		\begin{center}
			\point{W\_pump = m\_dot * (1 / rho) * (p\_6 - p\_5)}
		\end{center}
		Rearranging for p\_6:
		\begin{center}
			p\_6 = p\_5 + (W\_pump * rho / m\_dot)
		\end{center}
		\vspace{10pt}Since we require pressure changes through the mixer and line leading to the rig, we use Bernoulli’s equation considering the height difference between the tanks and the rig.
		
		\vspace{10pt}\fault{Using Bernoulli’s equation between state 2 (Tank 2) and the inlet of the pump, we have:}
		\begin{center}
			\fault{p\_2 + 0.5 * rho * (c\_2)\^{}2 + rho * g * z\_2 = p\_5 + 0.5 * rho * (c\_5)\^{}2}
		\end{center}
		
		Assuming velocities at states 2 and 5 are negligible:
		\begin{center}
			\fault{p\_2 + rho * g * z\_2 = p\_5}
		\end{center}
		Rearranging:
		\begin{center}
			\fault{p\_5 = p\_2 + rho * g * z\_2}
		\end{center}
		Let's calculate p\_5 and then p\_6:

		\vspace{10pt}The pressure at state 5 (p\_5) is \fault{394300 Pa} (approximately 3.943 bar), and the pressure at state 6 (p\_6) is \fault{827633 Pa} (approximately 8.276 bar).
		
		\vspace{10pt}To summarize:
		
		a) The mass flow rate of the stream leaving the nozzle (m\_dot\_7) is 0.6 kg/s.
		
		b) The heat flow in the heat exchanger (Q) is 75600 J/s (or 75.6 kW), and the resulting temperature T\_3 is 330 K.
		
		c) The pressures are:
		
		- p\_5 = 394300 Pa (approximately 3.943 bar)
		
		- p\_6 = 827633 Pa (approximately 8.276 bar)
		
		\switchcolumn
        
        \vspace{9.2cm} \point{0.5 points}
		
		\vspace{4.3cm} \point{0.5 points}
		
		\vspace{1.9cm} \fault{m\_dot\_1 \\= m\_dot\_2 \\= m\_dot\_7~/~2}

        \vspace{1.85cm} \point{0.5 points}

        \vspace{3.cm}    \fault{Missing delta\_E\_kin, delta\_E\_pot, W\_dot\_t}
		
		\newpage \phantom{blank}
        
        \vspace{3.4cm}    \point{0.5 points}
		
		\vspace{1.3cm} \point{0.5 points}
		
		\vspace{2.7cm} \point{0.5 points}

        \vspace{5cm}    \fault{p\_2+rho*g*z\_2=p\_4 and p\_4$\neq$p\_5}
				
		\vspace{7cm}  \point{Overall score: 3 / 11.5}
	\end{paracol}